\documentclass[aip,graphicx]{revtex4-1}

\usepackage{graphicx}
\usepackage{bm}
\usepackage{amssymb}
\usepackage{amsfonts}
\usepackage{amsmath}

\draft 

\newcommand{\Pe}{\mathrm{Pe}}
\newcommand{\Perel}{\mathrm{Pe}_{\mathrm{rel}}}
\newcommand{\kT}{k_{B}T}
\newcommand{\DkT}{D_{T}}
\newcommand{\Drel}{D_{\mathrm{rel}}}
\newcommand{\Dactive}{D_{A}}

\newcommand{\ent}{\mathrm{Ent}}
\newcommand{\cl}{\mathrm{Loop}}

\newcommand{\lmin}{\lambda_{\mathrm{min}}}

\newcommand{\Mob}{\bm{\mathcal{G}}}
\newcommand{\Mobtt}{\Mob^\mathrm{TT}}
\newcommand{\Mobtr}{\Mob^\mathrm{TR}}
\newcommand{\Mobrt}{\Mob^\mathrm{RT}}
\newcommand{\Mobrr}{\Mob^\mathrm{RR}}

\newcommand{\vv}[1]{\mathbf{#1}}
\newcommand{\disp}{\Delta}
\newcommand{\vecdisp}{\bm{\disp}}

\newcommand{\vecr}{\textbf{r}}
\newcommand{\vecF}{\textbf{F}}
\newcommand{\vecv}{\textbf{v}}

\newcommand{\vecomega}{\bm{\omega}}

\newcommand{\rhat}{\hat{\textbf{r}}}

\newcommand{\zhat}{\hat{\textbf{z}}}
\newcommand{\vhat}{\hat{\textbf{v}}}

\newcommand{\tumble}{\mathrm{tumble}}
\newcommand{\run}{\mathrm{run}}

\newcommand{\Ttum}{t_{\tumble}}
\newcommand{\Trun}{t_{\run}}

\newcommand{\Trot}{\tau_\mathrm{rot}}
\newcommand{\Trt}{\tau_\mathrm{r\& t}}

\newcommand{\Ttot}{\tau_\mathrm{tot}}
\newcommand{\Ts}{\tau_s}
\newcommand{\Tloop}{\tau_L}
\newcommand{\TPe}{\tau_B}
\newcommand{\TPeent}{\tau_{B,\ent}}
\newcommand{\TPeloop}{\tau_{B,\cl}}

\newcommand{\Cvv}{C_{vv}}
\newcommand{\Cvvp}{\Cvv^{(p)}}
\newcommand{\Cvvs}{\Cvv^{(s)}}

\begin{document}


\title{Influence of thermal fluctuations on active diffusion at large P\'{e}clet numbers} 



\author{O. T. Dyer}
\email[]{oliver.dyer@warwick.ac.uk}
\author{R. C. Ball}
\affiliation{Department of Physics, University of Warwick, Coventry, CV4 7AL, United Kingdom}

\date{\today}

\begin{abstract}
Wavelet Monte Carlo dynamics simulations are used to study the dynamics of passive particles in the presence of microswimmers, taking account of the often-omitted thermal motion alongside the hydrodynamic flows generated by the swimmers.
Although the P\'{e}clet numbers considered are large, we find the thermal motion to have a significant effect on the dynamics of our passive particles, and can be included as a decorrelation factor in the velocity autocorrelation with a decay time proportional to the P\'{e}clet number.
Similar decorrelation factors come from swimmer rotations, e.g.~run and tumble motion, and apply to both entrainment and far field loop contributions.
These decorrelation factors lead to active diffusivity having a weak apparent power law close to $\Pe^{0.2}$ for small tracer-like particles at P\'{e}clet numbers appropriate for \textit{E. coli} swimmers at room temperature.
Meanwhile, the reduced hydrodynamic response of large particles to nearby forces has a corresponding reduction in active diffusivity in that regime.
Together, they lead to a non-monotonic dependence of active diffusivity on particle size that can shed light on similar behaviour observed in experiments by Patteson et al.
\end{abstract}


\maketitle

\section{Introduction}

The influence of microswimmers, e.g.~algae or bacteria, on the dynamics of passive particles has received much attention since Wu and Libchaber first observed enhanced diffusion of spherical beads suspended in a soap film with \textit{E.~coli} \cite{Wu2000}.
The subsequent research into the swimmer-induced diffusion, often simply called `active diffusion', of colloids or infinitesimal tracer particles has spanned experiments \cite{Soni2003,Leptos2009,Mino2011,Kurtuldu2011,Valeriani2011,Mino2013,Jepson2013,Jeanneret2016,Mathijssen2018}, simulations \cite{Valeriani2011,Underhill2008,Molina2014,Morozov2014,Krafnick2015,Krishnamurthy2015,deGraaf2017,Shum2017,Harder2018} and analytic calculations \cite{Mino2013,Mathijssen2018,Dunkel2010,Thiffeault2010,Eckhardt2012,Mathijssen2015,Thiffeault2015,Suma2016,Burkholder2017,Yasuda2017,Mueller2017}, each in both 3 and (quasi-)2 dimensions.

While the precise results vary with the details of each system, all find the passive particles to exhibit enhanced motion over and above their own thermally-driven Brownian motion.
This swimmer-induced motion is super-diffusive on the time-scales of interactions with passing swimmers, and diffusive thereafter \cite{Mino2013,Krafnick2015}.

Until recently the size of the passive particles has received little attention, with a range of sizes used across the literature but typically a constant size within a given study.
Nevertheless one can identify several relevant properties that change with particle size: the thermal diffusivity and by extension the P\'{e}clet number $\Pe$ (defined as the ratio of advective and diffusive transport rates); the range of steric interactions with swimmers; and the near-field hydrodynamic response to swimmers \cite{Faxen1922,DurlofskyBradyBossis1987,RotnePrager1969,Yamakawa1970}.

The influence of P\'{e}clet number was included in a theoretical study by Kasyap et al.~\cite{Kasyap2014}, which predicts a peak in swimmer-induced diffusivities at moderate $\Pe$, rising from 0 at $\Pe=0$ (corresponding to infinitesimal tracers) and limiting to a finite value as $\Pe\rightarrow \infty$.
While this work accounted for particle size in the passive particles' thermal motion, they were hydrodynamically coupled to the swimmers by the (unregularised) Oseen tensor so that they were treated as point particles for hydrodynamic purposes.
Their results might therefore not be expected to quantitatively match real systems except when at small $\Pe$.

Shum and Yeomans have performed detailed boundary element simulations of single swimmer-passive interactions with a wide range of passive particle sizes, neglecting all thermal motion \cite{Shum2017}.
From these results they obtained the dilute limit active diffusivity by integrating over impact parameters.
In doing so they found a non-trivial dependence on the ratio of passive and active particle size, with a maximum active diffusivity for similarly sized particles when using a squirmer type swimmer model, but a minimum instead with bacteria models.

An experimental investigation of passive particle size active systems of \textit{E.~coli} was conducted by Patteson et al.~\cite{Patteson2016}, who found a non-monotonic variation in active diffusivity, peaking when the swimmer and passive particle sizes are similar.
This runs counter to the results for Shum and Yeomans' bacterial model.
This non-monotonicity suggests regimes where different particle sizes are dominated by different physics.
Since thermal diffusivity rises rapidly as particles get smaller, $D\sim R^{-1}$, we hypothesise this is important and will study simulations of analogous systems with this in mind.

It is notable that thermal fluctuations have usually been omitted from simulations of microswimmers on the basis that swimmer P\'{e}clet numbers are generally much greater than unity.
Work that has included thermal motion has done so in a hydrodynamically decoupled way \cite{Krishnamurthy2015}.
While this can capture some of the physics, the hydrodynamic coupling is required to provide the correct \emph{relative} thermal motion.

To elucidate the potential importance of thermal fluctuations we draw parallels to Taylor dispersion in pipes containing a steady shear flow \cite{Taylor1953,Mathijssen2018}.
In the absence of thermal fluctuations passive particles are carried parallel to the pipe axis at constant speed which is fastest at the centre of the pipe.
Thermal fluctuations allow particles to cross stream lines, with a corresponding change of advective velocity, leading to dispersion of particles in the stream direction.
Although the microswimmer flow fields are more complex, the crossing of stream lines is still expected to disperse advective motion, and will therefore have some influence on the active diffusion of passive particles, even if the Brownian motion itself is small.
In the slightly more limited context of particle entrainment, this effect has been seen to lead to a non-monotonic distribution of entrainment jump sizes with particle size \cite{Mathijssen2018}.

Another reason for the limited microswimmer work including thermal fluctuations is the great computational expense required to include the correct hydrodynamics of thermal fluctuations, which is a challenge well known to the polymer community where it has spawned a wide range of simulation algorithms \cite{Pham2009,Jain2012,DyerBall2017}.
Here we make use of the recently developed Wavelet Monte Carlo dynamics (WMCD) algorithm to include hydrodynamically coupled thermal fluctuations efficiently \cite{DyerBall2017,Dyer2019}.

After setting out the hydrodynamic theory and simulation details in Sections \ref{Sec: Theory} and \ref{Sec: Simulation details}, we will demonstrate the validity of using WMCD for active systems in Section \ref{Sec: Loop analysis}, where we use trajectories of passive particles in the flow of a swimmer on an infinite straight path as a test case.
We then study the dynamics of passive particles in dilute mixtures of swimmers, looking in detail at the role of particle size and temperature on the velocity autocorrelation and active diffusivity in Sections \ref{Sec: Cvv analysis} and \ref{Sec: Active diff results} respectively.
In doing so we find we can include the effect of all relevant time scales through exponential decay factors in the velocity autocorrelation, allowing us to construct an ansatz function that successfully captures the complex behaviour seen in the active diffusivity.


\section{Theory}
\label{Sec: Theory}

\subsection{Active diffusivity}

The total diffusivity of a particle can be split into thermal and active contributions as
\begin{equation}
D = \DkT + \Dactive.
\end{equation}
The thermal diffusivity for a sphere of radius $a$ in fluid of viscosity $\eta$ is given by the well-known Stokes-Einstein relation, which in a periodic cubic box of side length $L$ is corrected to

\begin{equation}
\DkT = \frac{\kT}{6\pi\eta a} \left( 1 - 2.837 \frac{a}{L} \right)
\end{equation}
to first order \cite{Dunweg1993}.

Rather than thermal fluctuations, the active diffusivity $\Dactive$ is driven by hydrodynamic and steric interactions with active particles in the system, and can itself be written as the sum of those contributions \cite{Pushkin2013PRL}.
To simplify data analysis we will not include steric interactions in this work so $\Dactive$ is purely hydrodynamic.

Regardless of its contributions, the active diffusivity can be expressed in terms of the velocity autocorrelation
\begin{equation}
C_{vv}(t) = \langle \vecv(t) \cdot \vecv(0) \rangle
\end{equation}
via the Green-Kubo relation \cite{Green1954,Kubo1957}
\begin{equation}
D_{A} = \frac{1}{3} \int\limits_{0^{+}}^{\infty} dt \,C_{vv}(t) .
\label{Eq: Green-Kubo relation}
\end{equation}
The lower limit $0^{+}$ denotes time $t\rightarrow 0$ from above, such that the thermal contribution is excluded when working in the overdamped limit (see the next section).
In practice this means the $t=0$ value we use in our data is extrapolated back from data at small but finite $t$.

\subsection{Equations of motion}

Working on time scales where thermal fluctuations can be considered instantaneous, or equivalently on time scales longer than the fluid relaxation time, leads to the overdamped Langevin equations for translational (superscript T) and rotational (R) velocities:
\begin{equation}
\vecv_{i} = \sum\limits_{j} \Mobtt_{ij} \cdot \vecF_{j} + \sum\limits_{j} \Mobtr_{ij} \cdot \bm{\Gamma}_{j} + \bm{\xi}_{i}
\label{Eq: Active Langevin}
\end{equation}
and
\begin{equation}
\vecomega_{i} = \sum\limits_{j} \Mobrt_{ij} \cdot \vecF_{j} + 
	\sum\limits_{j} \Mobrr_{ij} \cdot \bm{\Gamma}_{j} + \bm{\Xi}_{i},
\label{Eq: Active rotational Langevin}
\end{equation}
where $\vecF_{j}$ and $\bm{\Gamma}_{j}$ denote the force and torque at $\vecr_{j}$, which may or may not be located on a particle.
In this work $i$ will correspond to a particle, while the sum over $j$ corresponds to a sum over the swimmer forces described in Section \ref{Sec: swimmer model}.
We will not apply any point torques, so $\bm{\Gamma}_{j}=0$ for all $j$, but their inclusion here is useful for introducing the rotational mobility tensors which will be needed for correlations between the thermal fluctuations, $\bm{\xi}$ and $\bm{\Xi}$. 

The Lorentz reciprocal theorem links the TR and RT tensors by the transpose \cite{Happel1973}
\begin{eqnarray}
\Mobtr_{ij} & = & \left(\Mobrt_{ji} \right)^\textsf{T}.
\end{eqnarray}
For spheres, the unregularised versions of these tensors are to leading order in $1/r$:
\begin{eqnarray}
\Mobtt_{ij} & = & \frac{\delta_{ij}}{6\pi\eta a_i}\textbf{I} +  \frac{1-\delta_{ij}}{8\pi\eta r_{ij}} 
	\, ( \textbf{I} + \rhat_{ij}\otimes\rhat_{ij}) ,
\label{Eq: Standard Gtt tensor}\\
\Mobrt_{ij} & = & \delta_{ij}\, \textbf{0}  -  \frac{1-\delta_{ij}}{8\pi\eta r_{ij}^{2}} \, [\rhat_{ij}]_{\times},
\label{Eq: Standard Grt tensor}\\
\Mobrr_{ij} & = & \frac{\delta_{ij}}{8\pi\eta a_{i}^{3}} \textbf{I} + \frac{1-\delta_{ij}}{16\pi\eta r_{ij}^{3}} 
	\, ( 3\rhat_{ij}\otimes\rhat_{ij} - \textbf{I}),
\label{Eq: Standard Grr tensor}
\end{eqnarray}
where $a_{i}$ is the radius of particle $i$, $\delta_{ij}$ is the Kronecker-delta and $[\rhat]_{\times}$ is the skew-symmetric tensor expressed as $\varepsilon_{abc}\hat{r}_{b}$ in index notation.

The mobility tensors used in this work are those that appear in Wavelet Monte Carlo Dynamics (WMCD) \cite{DyerBall2017,Dyer2019}, described below, which smoothly bridge the large $r$ and $\delta_{ij}$ terms.
Although reached in a completely different way, they closely approximate the tensors obtained by using Fax\'{e}n's laws to sum the fluid flow over the particle surface \cite{DurlofskyBradyBossis1987}.

Finally, the fluctuation dissipation theorem gives \cite{Kubo1966,Noetinger1990}
\begin{eqnarray}
\left\langle \bm{\xi}_{i}(t) \otimes \bm{\xi}_{j}(t') \right\rangle & = & 2 \kT\, \Mobtt_{ij}\, \delta(t-t'), 
\label{Eq: tt noise correlations}\\
\left\langle \bm{\Xi}_{i}(t) \otimes \bm{\xi}_{j}(t') \right\rangle & = & 2 \kT\, \Mobrt_{ij}\, \delta(t-t'), 
\label{Eq: tr noise correlations}\\
\left\langle \bm{\Xi}_{i}(t) \otimes \bm{\Xi}_{j}(t') \right\rangle & = & 2 \kT\, \Mobrr_{ij}\, \delta(t-t')
\label{Eq: rr noise correlations}
\end{eqnarray}
as the correlations between noise terms.


\section{Simulation details}
\label{Sec: Simulation details}

\subsection{Wavelet Monte Carlo dynamics}

We use a smart WMCD simulation algorithm, for which the full details can be found in Ref.~\citenum{Dyer2019}.
Only the physically important details are listed here.

Systems in smart WMCD are evolved as per Eqs.~\eqref{Eq: Active Langevin} and \eqref{Eq: Active rotational Langevin} using a sequence of wavelet and plane wave Monte Carlo moves, which can displace between 1 and all particles depending on the chosen move parameters.
As per smart Monte Carlo \cite{Rossky1978}, the bias on the move parameters supplies the causal terms, while the variance in parameters supplies the thermal fluctuations.

The hydrodynamic interactions, i.e.~the mobility tensors, arise implicitly by careful choice of parameter distributions and the possibility of single moves containing multiple particles.
The smooth approach to the $r=0$ hydrodynamic tensors is achieved by setting finite values for minimum wavelet radii $\lmin$, which is chosen separately for both translations and rotations at each particle size to give the appropriate particle mobility at $r=0$.

The end result is an efficient algorithm that includes long ranged hydrodynamic correlations for both causal and thermal forces with a computational cost that rises with the total number of particles as $N\ln N$ per unit of physical time.
The price is that it is limited to the mobility tensors for spheres and cannot currently handle lubrication forces or no-slip boundary conditions on the sphere surface.
Nevertheless, the efficient inclusion of hydrodynamically coupled thermal fluctuations means WMCD is well placed to investigate whether these play a role in active-passive mixtures.

\subsection{Swimmer model}
\label{Sec: swimmer model}

\begin{figure}
    \includegraphics[width=8.5cm]{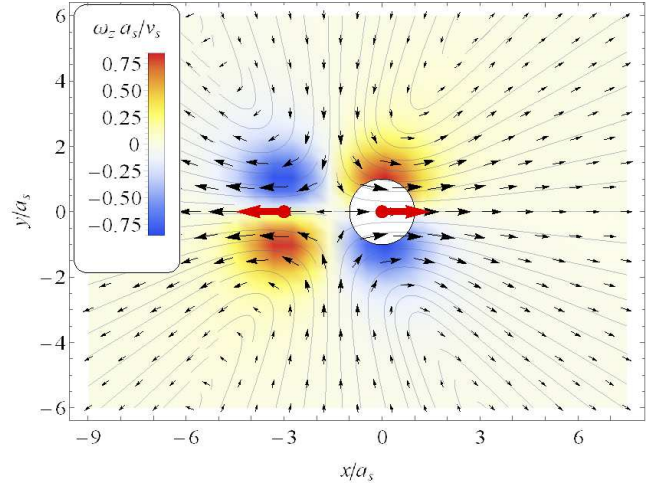}
    \caption{Schematic view of the 2-force model for a pusher type swimmer and its associated flow fields, as generated in WMCD.
    The swimmer's body is indicated by the white disk, which moves in the positive $x$-direction as per the translational flow field (black arrows) at its centre.
    Red arrows indicate the position and direction of the swimming forces, while the coloured background shows the rotation field which is always into/out of the plane with negative/positive $\omega_{z}$ respectively.
    The full 3-dimensional flow field is symmetric under rotations about the $x$-axis.
    Field strengths are in arbitrary units.}
    \label{Fig: Swimmer flow fields}
\end{figure}

Our swimmers are represented by a simple two-force pusher-type model, with a forward force $\vecF_s$ placed at the swimmer centre and a tail force $-\vecF_s$ placed $A_s=3a_s$ behind the swimmer, as depicted in Fig.~\ref{Fig: Swimmer flow fields}.
Using the WMCD mobility tensor these produce a swimming velocity of
\begin{equation}
\vecv_{s} \approx 0.5 \frac{1}{6\pi\eta a_{s}} \vecF_{s}.
\end{equation} 
Encounters of passive and swimmer particles are governed by competition between the above swimming and the relative thermal diffusivity characterised by its large separation value 
\begin{equation}
    \Drel=\frac{k_B T}{6 \pi \eta}(1/a_s+1/a_p).
\end{equation}
We can define a corresponding P\'{e}clet number by using $a_s$ as the relevant length scale, and comparing the advection rate $v_{s}/a_{s}$ to the diffusion rate $\Drel/a_{s}^{2}$, leading to
\begin{equation}
    \Perel = \frac{a_{s} v_{s}}{\Drel}=\frac{\Pe_s}{1+a_s/a_p} 
\label{Eq: Peclet number}
\end{equation}
where $\Pe_s \approx 0.5F_{s} a_{s}/\kT$ is the intrinsic P\'{e}clet number of the swimmer alone.

The run and tumble motion exhibited by many micro-organisms, such as \textit{E.~coli}, is characterised by alternating phases of swimming and stopping, with increased rotational motion during the stopped `tumble' phase.
This has previously been modelled in simulations by instantaneous and random reorientations of swimmers at a Poisson distributed frequency \cite{Krishnamurthy2015,Berg1993}.

In this work the tumble phase is stretched out over a finite time $\Ttum$, during which there is an increased rotational diffusion and the swimming forces are turned off, so that the swimmer temporarily becomes a passive particle.
To reduce the number of variables and make the data easier to decipher, we use fixed values of $\Trun$ and $\Ttum$ rather than choosing them from a Poisson distribution.
Hence each swimmer in the system cycles between running and tumbling with the same period $\Trun + \Ttum$, although each is initialised at a different point in this cycle.

\subsection{Simulation parameters}

Table \ref{Table: System parameters} lists the values of the physical parameters used for the results sections.

\begin{table}
\caption{\label{Table: System parameters}System parameters in the results sections. }
\begin{ruledtabular}
\begin{tabular}{ccc}
Parameter & Section \ref{Sec: Loop analysis} & Sections \ref{Sec: Cvv analysis} \& \ref{Sec: Active diff results} \\
\hline
$L$ & $\infty$ &  40 $\mu$m \\ 
$N_{s}$ & 1 &  64\\ 
$a_s$ & 1 $\mu$m & 1 $\mu$m\\
$a_p$ & 0.5-2 $\mu$m &  0.125-2 $\mu$m \\ 
$T$ &  0.3 mK &  30-3000 K.\\ 
$v_s$ & 40 $\mu\mathrm{m}\,\mathrm{s}^{-1}$ & 40 $\mu\mathrm{m}\,\mathrm{s}^{-1}$ \\
$\eta$ & $0.85$ mPa\,s & $0.85$ mPa\,s\\
$\Pe_\mathrm{rel}$ & $0.5$-$1\times10^{8}$ & 2-1000 \\ 
$t_{\mathrm{run}}$ & $\infty$ & 1 s\\ 
$t_\tumble$ & n/a & 0.1 s\\
$\theta_\tumble$ & n/a & $70^\circ$
\end{tabular}
\end{ruledtabular}
\end{table}

Swimmer parameters are appropriate for \textit{E.~coli} \cite{Berg1993}, with only the run and tumble behaviour changing between sections.
Where tumbling is present, the distance swum between tumbles compared to the swimmer radius is
\begin{equation}
\lambda = v_s \Trun /a_s,
\end{equation}
equalling 40 in our simulations.
These runs dominate the time elapsed, so the active diffusion of the swimmers is well approximated by $ D_{A}^{(s)}\approx a_s v_s  \lambda / 6$.

Because we have a finite $\Ttum$ we need to specify how much swimmer orientations decorrelate when tumbling.
This is done via \cite{Saragosti2012}
\begin{equation}
    \langle \vhat(\Ttum) \cdot \vhat(0) \rangle = \exp(-2D_\tumble^{RR} t_\tumble),
\label{Eq: tumble angle}
\end{equation}
which will be useful for relating tumble angles to decorrelation times in Section \ref{Sec: Cvv analysis}.
In our simulations we used $\theta_\tumble = \arccos \langle \vhat(\Ttum) \cdot \vhat(0) \rangle  = 70^\circ$, inside the range of angles identified in Ref.~\citenum{Berg1993}.
Reorientation from tumbling and normal rotational diffusion are comparable when averaged over the run and tumble cycle, with tumbling being dominant and sub-dominant at the lower and higher temperatures respectively in Sections \ref{Sec: Cvv analysis} and \ref{Sec: Active diff results}.
We note that the small size of \textit{E. coli} is important here, and thermal rotations would be less significant were we modelling a larger microswimmer.

The swimmer volume fraction in Sections \ref{Sec: Cvv analysis} \& \ref{Sec: Active diff results} is $\phi_s=(4\pi/3) (a_s/L)^3  N_s \approx 0.42\%$.
The volume fraction of passive particles is unimportant because they do not influence each other's motion, despite having correlated displacements.
However, what is important for good statistics is the product of $N_p$ and total data collection time, which was a minimum of $1.85\times 10^{5}\Trun$ of effective single particle tracking time per data point.


\section{Low noise particle trajectories}
\label{Sec: Loop analysis}

Our first results focus on interactions between individual swimmers and passive particles in the idealised scenario where swimmers move along an infinite straight path at constant speed.
Such simulations have been done with more sophisticated techniques previously \cite{deGraaf2017,Shum2017,Dunkel2010,Pushkin2013PRL,Pushkin2013JFluidMech}, but are revisited here because they validate the use of WMCD for active systems while helping to visualise behaviour quantified in later sections.

Swimming along a perfectly straight path is not possible in WMCD, but a good approximation was achieved by switching off all rotations and reducing the temperature to raise the swimmer P\'{e}clet number to $1.3\times 10^{8}$.  
This also makes the P\'eclet number of relative motion high enough that thermal diffusion has negligible role in particle encounters.

\begin{figure}
    \includegraphics[width=7cm]{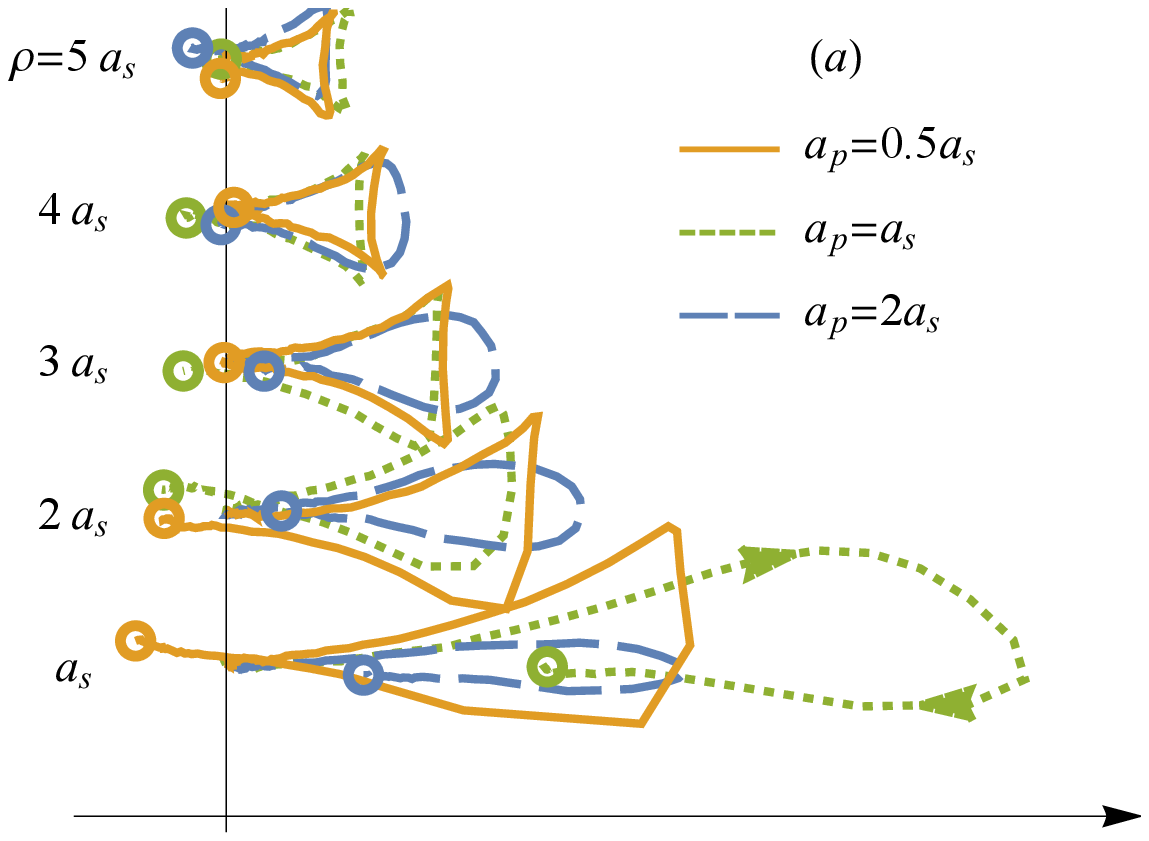}
    \centering
    \includegraphics[width=8.5cm]{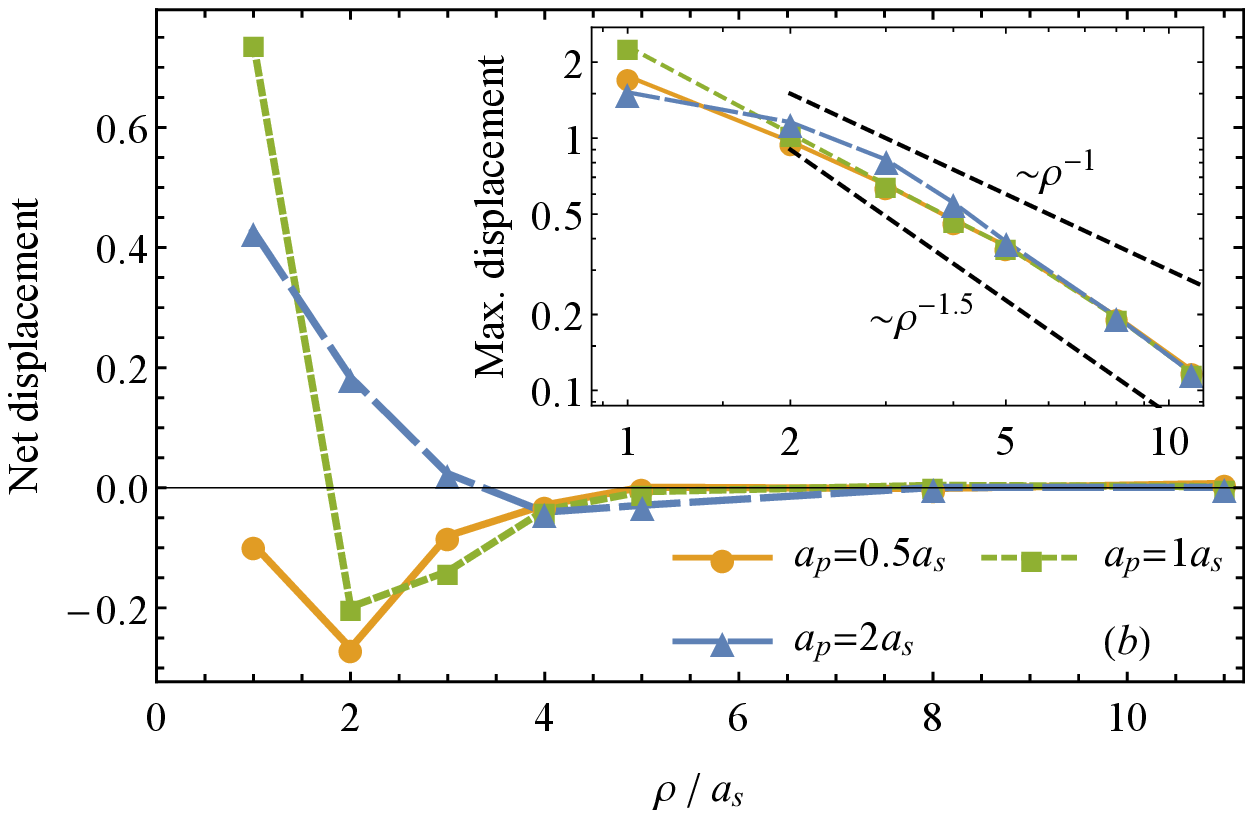}
    \caption{
    (a): Example trajectories of passive particles at different impact parameters with a swimmer on a straight path from left to right.
    All trajectories are travelled in a clockwise sense, as indicated on the $a_p=a_s,\; \rho = a_s$ trajectory, starting on the vertical line and ending at the rings capping each trajectory.
    (b): Plots of the net displacement over the whole trajectory, and the maximum displacement in the $x$-direction (inset).
    All data were averaged over 100 trajectories to smooth out thermal fluctuations.}
    \label{Fig: Particle loops}
\end{figure}

In these simulations, performed in an infinite box, a single swimmer was set swimming in a straight line between $-40a_{s}\hat{\textbf{x}}$ and $+40a_{s}\hat{\textbf{x}}$.
A single passive particle was placed an impact parameter $\rho$ off the swimmer's path at $\vecr_{p}(t=0)=\rho\hat{\textbf{y}}$, and its position was traced out as the swimmer passed by.
The $x$-$y$ components of these trajectories are plotted in Fig.~\ref{Fig: Particle loops}(a) for various $\rho$.

\begin{figure}
    \includegraphics[width=8.5cm]{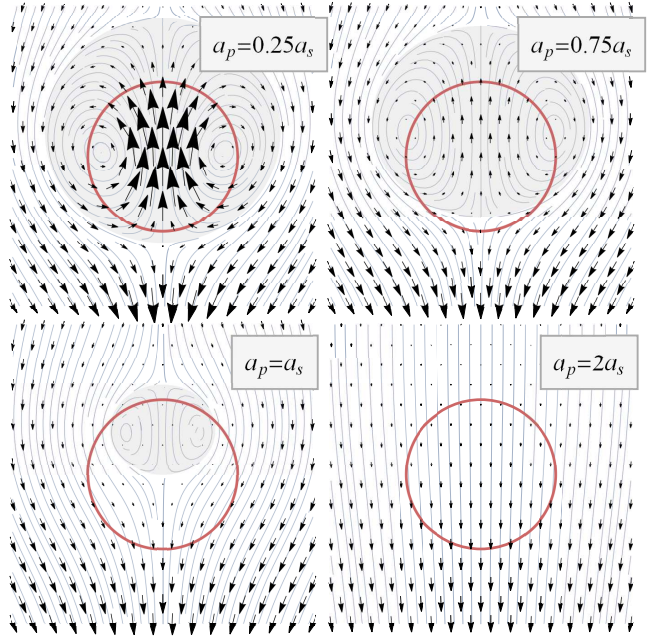}
    \caption{Translational flow fields relative to the swimmer velocity (upwards) as seen by particles of different sizes.
    Vector lengths across all plots have the same (arbitrary) units and can be compared directly.
    The disk, with radius $a_s$, indicates the swimmer size, and the flow field everywhere outside the plotted region is downwards.}
    \label{Fig: Near-field relative flows}
\end{figure}

Qualitatively, these trajectories match expectations by forming (almost) closed loops with cusps \cite{Dunkel2010} at large $\rho$, whilst at small $\rho$ the cusps become more rounded and the loops open up with a significant finite net displacement.
The details of how the loops change at small $\rho$ are sensitive to the near-field details of the hydrodynamic mobility tensors and hence to passive particle radius $a_p$, as evidenced by the clear differences between the loops for $a_p = a_s$ and $2a_s$ at same $\rho$. 
To demonstrate this sensitivity to particle size, Fig.~\ref{Fig: Near-field relative flows} shows the flow fields experienced by passive particles of different sizes, as seen in the swimmer's reference frame.
The most notable feature is the shaded recirculation zone (closed stream lines) close to the swimmer body that appears for small $a_p$, which the passive particles do not enter as the swimmer passes by.
The $a_p = 0.25a_s$ tile is very close to the flow-field that would be produced using the unmodified Oseen tensor, while the differences in the other 3 tiles arise due to the near-field corrections in the WMCD tensor.
These corrections are therefore responsible for the differences in the trajectories in Fig.~\ref{Fig: Particle loops}(a).
This highlights the limitations of treating passive particles as infinitesimal tracers when near-field flows are important.

Fig.~\ref{Fig: Particle loops}(b) provides a more quantitative description of the loops by plotting their size as measured by their maximal displacement parallel to the swimmer's path.
The unvarying loop shape at large $\rho$ means this choice is equivalent to the different measure used by Shum and Yeomans \cite{Shum2017}.
Indeed, the decrease in loops size with a power law between $\rho^{-1}$ and $\rho^{-1.5}$ is consistent with their results.

It is the total net displacements of the passive particles that are key to their induced active diffusion, and these show a difference of sign and magnitude between large and small $\rho$.  
The large $\rho$ loops have negative net parallel displacement corresponding to a back-flow effect, and this can be estimated theoretically (see Appendix \ref{Sec: Loop size appendix} and Ref.~\citenum{Pushkin2013JFluidMech}).
The two key lengths here are $a_p$ and the distance between the centres of thrust and drag for our swimmer $A_s=3a_s$.  
When $\rho>a_p,A_s$ we can work from the far field flow of a force dipole leading to the estimate that $\disp (\rho) \sim - \rho^{-3}$. 

The trajectories for smaller $\rho$ have positive net parallel displacements. 
For the limited but relevant range $a_p<\rho<A_s$ we obtain theoretically the much weaker dependence $\disp(\rho) \sim + \rho^{-1}$ by treating the swimmer as two explicit point forces. 
In practice the crossover from positive to negative displacements can be seen in Fig.~\ref{Fig: Particle loops}(a) to be sensitive to the values $a_p/a_s=0.5,1,2$ investigated, with larger $a_p$ changing sign at larger $\rho$, but then having a smaller net displacement:  both effects are consistent with larger passive particles tracking a wider scale average of the advecting fluid flow.

So far we have only considered trajectories with a well defined net displacement in the reference frame of the background fluid.
These correspond to moving along the open stream lines in Fig.~\ref{Fig: Near-field relative flows}.
The trajectories of particles inside the shaded recirculation zone around the swimmer are instead well defined in the swimmer's reference frame: they have zero net displacement in this frame.
The passive particle is therefore displaced by $v_s t$ in the fluid frame, if trapped in the recirculation zone for time $t$.
This behaviour is akin to entrainment observed in real systems \cite{Jeanneret2016,Mathijssen2018}, albeit driven by internal flows rather than steric repulsion and a non-slip boundary at the swimmer surface.
We therefore describe such trajectories as `entrained' in the following sections.


\section{Velocity Autocorrelations}
\label{Sec: Cvv analysis}

\begin{figure}
    \includegraphics[width=8.5cm]{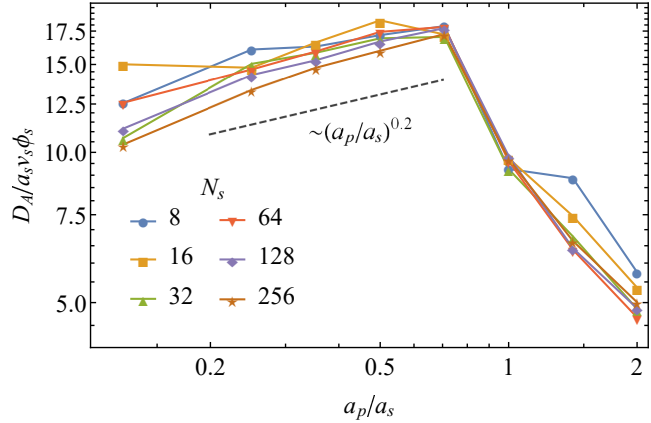}
    \caption{Plots of active diffusion of passive particles of different sizes, in systems with different swimmer volume fractions, $\phi_s \propto N_s$. 
    The vertical axis is scaled by $\phi_s$, leading to the collapse of the data sets, thereby confirming $D_A \propto \phi_s$.
    The dashed line gives a rough guide to the weak power law behaviour seen on the small $a_p/a_s$ side of the data.}
    \label{Fig: Dactive propto phi_s}
\end{figure}

Next we discuss the velocity autocorrelation of passive particles in active systems.
The systems are now made periodic with cubic box, we use a reference temperature $T_0$ representative of 300K, and swimmers are free to rotate by rotational diffusion, hydrodynamic interactions and tumbling.
Note that in these simulations we confirmed $\Dactive$ is proportional to swimmer concentration, as demonstrated in Fig.~\ref{Fig: Dactive propto phi_s} whose results are discussed in detail in Section \ref{Sec: Active diff results}, so we are in the dilute limit where swimmer-swimmer interactions can be neglected. 
We also note that we find $D_A$ to vary roughly as $(a_p/a_s)^{0.2}$ at small $a_p$. 
This power cannot be explained by considering any one mechanism, and is a sign that we are in a complex regime with many contributing factors.
$D_A$ itself has integrated out too much information to unpick these factors, motivating our focus on velocity autocorrelations instead.

\begin{figure}
    \includegraphics[width=8.5cm]{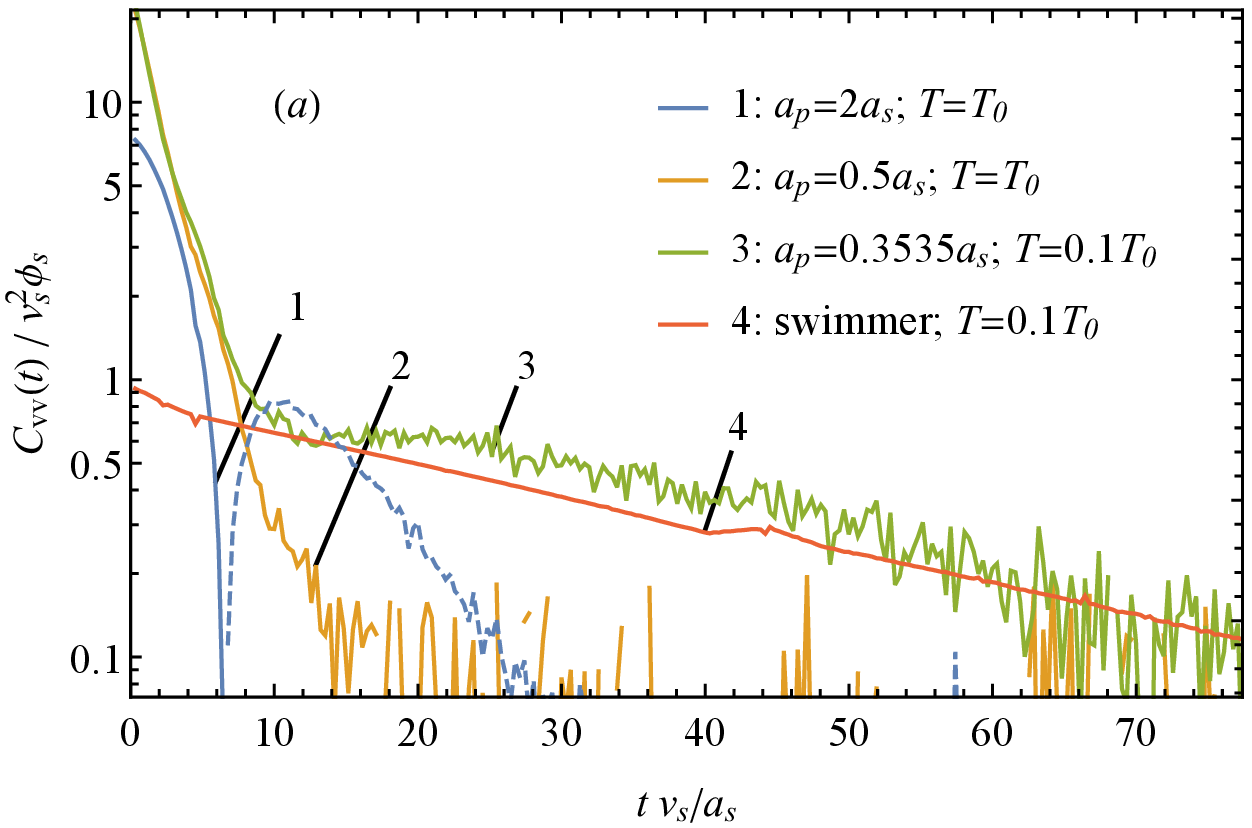}
    \includegraphics[width=8.5cm]{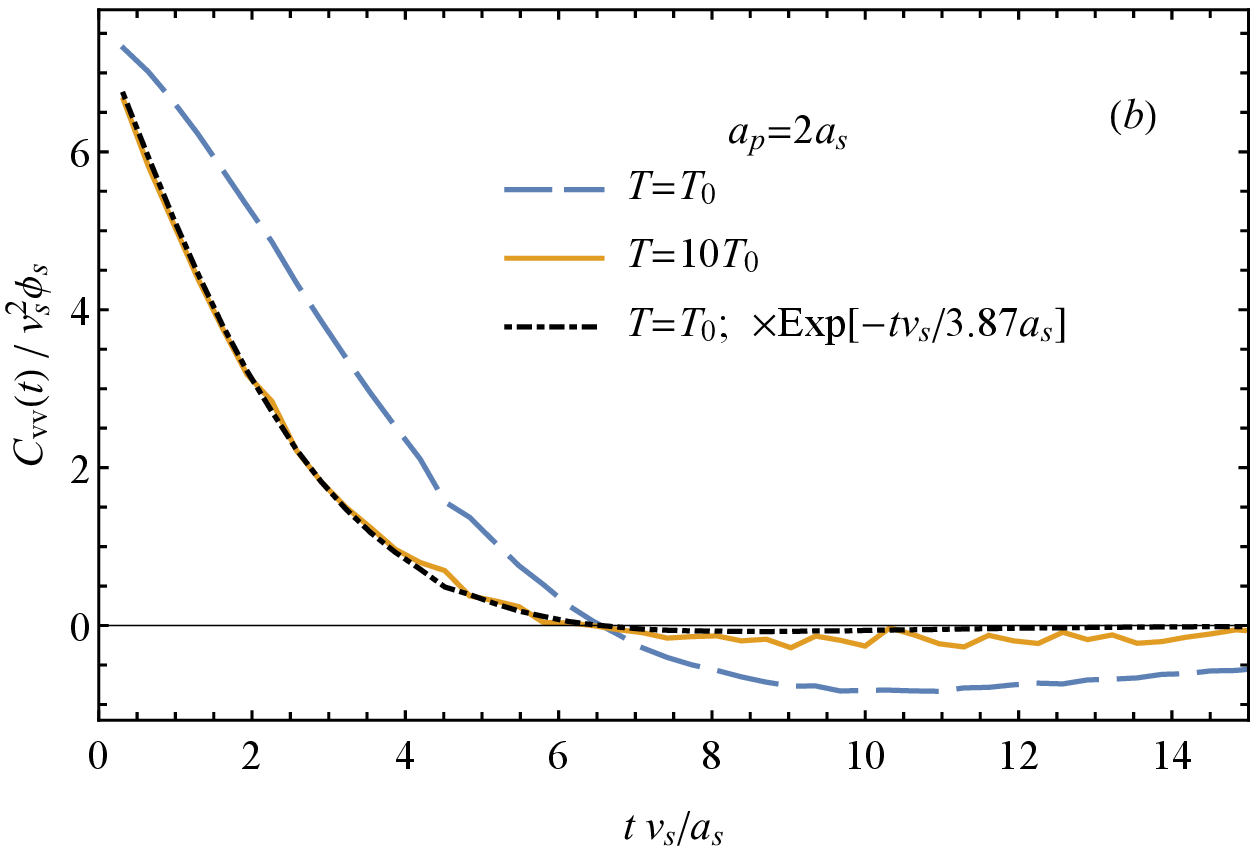}
    \caption{Example velocity autocorrelations exhibiting the distinct forms observed across all data collected.
    (a): data shown on a linear-log scale, with passive particle sizes and temperatures as indicated on the legend.
    The swimmer data (4) includes a factor of $\phi_s$ relative to the passive particle data (1-3) so that it starts at $\Cvv(0)/v_s^2 \approx 1$.
    The dashed line is a continuation of curve 1 with a negative sign to see it on the log scale.
    (b): re-plots curve 1 with a linear vertical scale alongside data for the same $a_p$ at a higher temperature.
    The dot-dashed line demonstrates there is an exponential decay factor between the two curves.}
    \label{Fig: Example Cvv}
\end{figure}

We begin with a qualitative discussion of these and will use this as the basis for a quantitative discussion in Section \ref{Sec: Ansatz function}.
We will denote passive particle and swimmer autocorrelations with $\Cvvp$ and $\Cvvs$ respectively.
Fig.~\ref{Fig: Example Cvv}(a) shows the three forms of $\Cvvp$ we find in our data, alongside an example $\Cvvs$ which is a simple exponential decay with some fine details coming from run and tumble motion which we will ignore.

In curve 1, which is typical of systems with $a_p \geqslant a_s$, we see negative tails compatible with the forwards-backwards movement in the loop trajectories discussed in the preceding section. 
In curve 2, with smaller $a_p$ at the same temperature, this negative tail appears to have vanished, or at least been reduced to the size of noise in our data.

Note the value of $\Cvvp(0)$ has increased between these curves, which is due to small particles having a larger response to the swimming forces in the near-field.
The size of the increase, at more than a factor of 2, is indicative of how much the near-field contributes to $\Cvvp(0)$, and we anticipate a similar rise in $D_A$.
Importantly, however, $\Cvvp(0)$ reaches a maximum at around $a_p=0.5a_s$, below which it is essentially constant.
This will be discussed in greater detail in Section \ref{Sec: Ansatz loop term}.

Finally, curve 3 in Fig.~\ref{Fig: Example Cvv}(a) shows that the reduction of temperature reveals a long-time exponential tail.
This tail is present in the all our $T=0.1T_0$ data with $a_p\leqslant a_s/\sqrt{2}$, although its amplitude is not always the same.
This tail runs almost parallel to $\Cvvs$ in curve 4, suggesting the passive particles are tracking the swimmer motion, that is they are entrained.
This is further supported by the quantitative similarity between $\Cvvp(t)$ and $\phi_s \Cvvs(t)$, which is expected if a fraction of order $\phi_s$ of passive particles are entrained at any one time.
This entrainment is both initiated and ended by crossing the boundary of the closed stream lines.

There is in fact a subtle but important difference in the gradients of curves 3 and 4 at large times.
First, tracing the long-time exponential back to $t=0$ leads to a value larger than $\phi_s \Cvv^{(s)}(0)$, fitting with the entrainment volume around a swimmer being larger than the swimmer's own volume, as per Fig.~\ref{Fig: Near-field relative flows}.
Second, this means the mid-time section of the curve undershoots the long-time exponential, which is consistent with there being a negative contribution akin to that seen in curve 1.
This too is expected as the loop trajectories should still be present, and indeed the location of the peak of the negative (dashed) section of curve 1 coincides with the depression in curve 3.
We believe this story applies to curve 2 as well, but in this case the negative contribution is closely matched to the positive tail so they cancel each other out.

Our final comment on the qualitative features of $\Cvvp$ is that we find curves at different temperatures but the same particle size can be successfully mapped onto each other by introducing an exponential decay factor between them.
This is demonstrated in Fig.~\ref{Fig: Example Cvv}(b), and will be central to our approach going forwards.

\subsection{Ansatz function}
\label{Sec: Ansatz function}

We now attempt to quantify the velocity autocorrelations, starting with the swimmers since they have the simplest form, and we have seen that the passive particles can pick up the same form.

Ignoring some fine details seen in Fig.~\ref{Fig: Example Cvv}(a) that arise due to the fixed times in the run and tumble cycle, it is clear we have a simple exponential decay of the form,
\begin{equation}
    C_{vv}^{(s)}(t) = v_{s}^{2} \frac{\Trun}{\Trun+\Ttum}
        e^{-t/\Ts}
\label{Eq: Swimmer Cvv}
\end{equation}
where the run and tumble factors account for the time spent not swimming and hence why the swimmer curve begins just below 1.
The exponential decay comes only from reorientations because translational diffusion does not change the swimming direction, and therefore does not affect $C_{vv}^{(s)}$ beyond the Brownian spike at $t=0$, which we are ignoring.
Hence we have
\begin{equation}
    \Ts^{-1} = \Trt^{-1} + \Trot^{-1}
\end{equation}
where the decay time associated with normal rotational diffusion is 
\begin{equation}
    \Trot = \left( 2 D_{s}^{RR} \right)^{-1} = (2/3) \Pe_s a_s /  v_s  
\label{Eq: rot diffusion time}
\end{equation}
and the decay time for run and tumble motion, spread across the whole run and tumble cycle, is
\begin{equation}
\Trt = \left( 2 D_{\mathrm{tum}}^{RR}\frac{\Ttum}{\Trun + \Ttum} \right)^{-1} = 
    1.02\Trun.
\end{equation}
Note the final expression uses Eq.~\eqref{Eq: tumble angle} and our typical tumble angle of $70^\circ$.
Together, these predict $\Ts v_s / a_s = 39.2$, in good agreement with the decay time observed in curve 4 in Fig.~\ref{Fig: Example Cvv}(a).

We also expect the $\Cvvp$ to decay with $\Ts$.
In the entrainment tail the reasoning is the same as for the swimmers - the decorrelation of the direction of $\vecv(t)$ - but we also expect it to play a role in decorrelating non-entrained loops where swimmer rotations effectively force the passive particles onto different stream lines, as well as rotating the flow field.
By similar reasoning, we expect an additional decorrelation time coming from (translational) Brownian motion, $\TPe$.
This should also feature in the entrainment tail where it drives entry and escape of the entrained volume.
In higher density systems we might expect a similar term for swimmer-swimmer interactions, but we do not consider those here.

We express the total $\Cvvp$ as the sum of terms from entrainment and non-entrained loops:
\begin{equation}
    \Cvv^{(p)}(t) = \Cvv^{(\ent)}(t) + \Cvv^{(\cl)}(t)
\label{Eq: Cvv ent plus loop}
\end{equation}
which we now detail separately.

\subsubsection{The entrainment term}

Entrained particles follow swimmers, therefore we expect $\Cvv^{(\ent)}$ to be given by Eq.~\eqref{Eq: Swimmer Cvv} modified to account for Brownian escape and the actual entrained volume fraction $\phi_\ent$. 
This implies that 
\begin{equation}
    \Cvv^{(\ent)}(t) = \phi_\ent \Cvv^{(s)}(t)\, \exp\left(-\sqrt{t/\TPeent}\right).
\end{equation}
The details of the decorrelation factor, including the perhaps unexpected square root, will be discussed after $\phi_\ent$.

As Fig.~\ref{Fig: Near-field relative flows} showed, $\phi_\ent$ varies with particle size, and has 3 regimes of behaviour.
For $a_p > a_s$, there is no entrainment as the response to the swimming forces is always less than that of the swimmer.
For $a_p \ll a_s$ the entrainment volume is constant and is approximately 4 times the swimmer volume for our model.
As $a_p$ increases the volume begins to decrease when the hydrodynamic near-field of the particle, i.e~the distance within which the hydrodynamic response is different to the Oseen tensor, is comparable to the geometric size of the swimmer, $A_s$.
In our WMCD simulations that hydrodynamic range is $5.35a_p$, leading us to define the ratio
\begin{equation}
    \mathcal{R} = \frac{5.35a_p}{A_s} = 1.78 \frac{a_p}{a_s}
\end{equation}
with which we would anticipate the regime change at $\mathcal{R} \approx 1$.
Assuming a simple linear interpolation between the large and small $a_p$ regimes, we have
\begin{eqnarray}
    \phi_\ent \approx \phi_s \left\lbrace      
        \begin{array}{ll}
            4 & \mathcal{R} < 1 \\
            0 & a_p > a_s \\
            9.12-5.12\mathcal{R} & \mathrm{between}
        \end{array}
    \right. .
\label{Eq: c_ent}
\end{eqnarray}

We now address the Brownian decorrelation factor.
This can be estimated by the fraction of particles left inside a stagnant, entrained sphere after diffusing for time $t$ if we assume an initially uniform distribution and that they are swept away, without return, upon first passage across the boundary of the entrained volume.
This calculation is the same as the one leading to Eq.~(6.19) in Ref.~\citenum{Crank1975}, which decays approximately as $\exp[-\sqrt{t}]$ at small times.
It is the asymmetry in this first passage problem that leads to the square root rather than the simple exponential decay used in other decay factors.

That calculation gives us a handle on the form of the decay time $\TPeent \sim a_s^2 / \Drel = (a_s/v_s)\Perel$.
Using all the other parameters in $\Cvv^{(\ent)}$ as described above, we set the numerical prefactor to match the entrainment tail in Fig.~\ref{Fig: Example Cvv}(a) curve 3.
This gives us
\begin{equation}
    \TPeent \approx 0.07 \Perel a_s / v_s, 
\end{equation}
corresponding to diffusing a distance of $\sqrt{6 \Drel\TPeent} \approx 0.65 a_s $, which is reassuringly less than $a_s$.

\subsubsection{The loop term}
\label{Sec: Ansatz loop term}

The non-entrained loop contribution in Eq.~\eqref{Eq: Cvv ent plus loop} needs to provide the negative tail seen in the $a_p=2a_s$ data in Fig.~\ref{Fig: Example Cvv}.
The functional form of this could in principle be calculated in the far-field where the loop trajectory is known mathematically for a dipole swimmer \cite{Dunkel2010}, but the near-field is not entirely absent in our expression so we would expect model-dependent terms to enter.
We therefore choose to take a simpler route and use a functional form that has the correct features:
\begin{multline}
    \Cvv^{(\cl)}(t)  = c_0 (1 - c_2 (t/\Tloop)^2)  \\ \times
        \exp\left[-\left(\Tloop^{-1} + \Ts^{-1} + \TPeloop^{-1} \right) t \right].
\label{Eq: Ansatz loop term}
\end{multline}
The task is now to identify all the new parameters, starting with the coefficients $c_0$ and $c_2$.

$c_2$ will not be explored in detail, but we note it must satisfy $0 < c_2 \leqslant 1/2$ to ensure both a negative tail exists and $\Cvv^{(\cl)}$ has a positive integral.
In Section \ref{Sec: Active diff results} we will use two values to show its effect on $D_A$.

\begin{figure}
    \includegraphics[width=8.5cm]{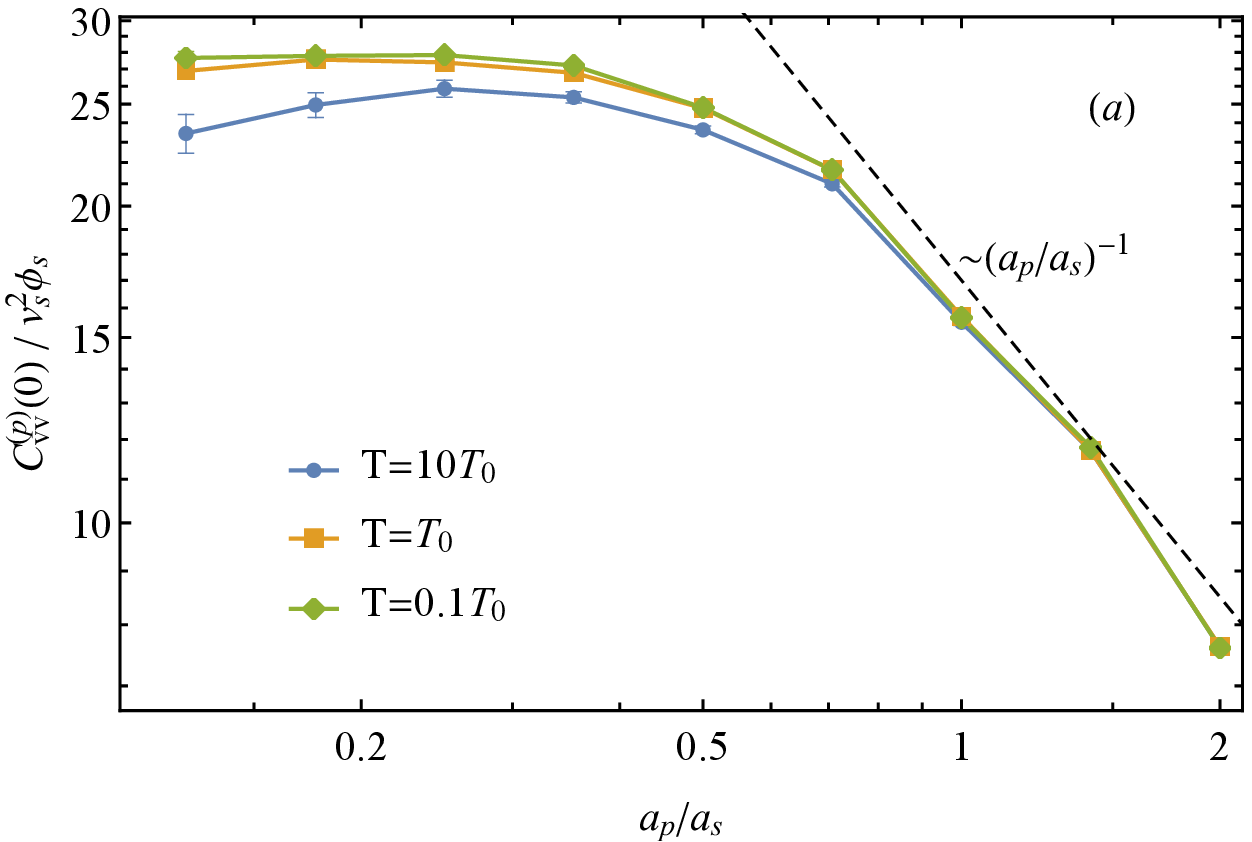}
    \includegraphics[width=8.5cm]{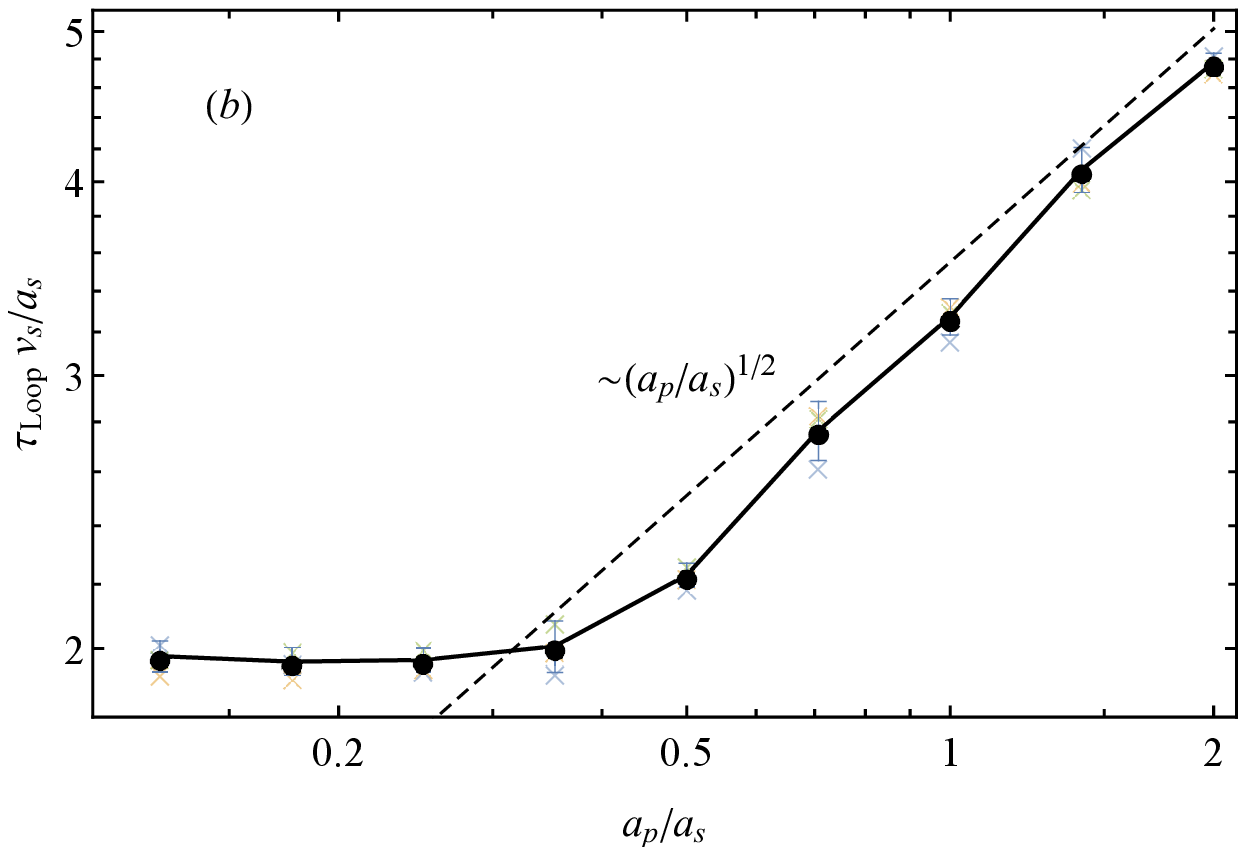}
    \caption{(a): Plot of $\Cvvp(0)$ against $a_p$ for our 3 different temperatures.
    (b): Plot of the ansatz parameter $\Tloop$ against $a_p$, with each marker showing the mean of the values for the 3 temperatures, whose individual values are marked with crosses at low opacity.
    The dashed guidelines in both plots are only to indicate the rough behaviour.}
    \label{Fig: anasatz params}
\end{figure}

$c_0$ sets the value at $t=0$ and can be written as
\begin{equation}
    c_0 = \Cvv^{(p)}(0) - \phi_\ent \Cvv^{(s)}(0).
\end{equation}
The total $\Cvv^{(p)}(0)$ is most easy to access in our data by interpolating small $t$ values back to 0 assuming a simple exponential decay, thereby avoiding the Brownian spike which overwhelms the zero-time data.
Values obtained by this procedure are shown in Fig.~\ref{Fig: anasatz params}(a), where we see 2 distinct regimes: $\Cvv^{(p)}$ is flat at small $a_p/a_s$ and decays at larger values.

The flat regime is simply the result of passive particles being small compared to the distance between the swimming forces, $A_s$, so they act like infinitesimal tracer particles.
Indeed, the regime change occurs close to $\mathcal{R}=1 \Leftrightarrow a_p/a_s = 0.56$, supporting this picture.
What is less easy to understand is the apparent dependence on temperature in this regime, with the high temperature data being too small to be accounted for by our error margins.
We attribute this apparent $T$-dependence to the assumption of a simple exponential decay when interpolating back, which underestimates the contribution from $\exp[-\sqrt{t/\TPeent}]$.
The corresponding error is largest at small $\Perel$ and only when the entrainment term is present, both fitting with where the difference occurs in Fig.~\ref{Fig: anasatz params}(a).

The decay at larger $a_p/a_s$ can be understood using scaling arguments, detailed in Appendix \ref{Sec: Scaling of Cvv and Tloop}, which use the fact our mobility tensor can be written as $a_p^{-1} \Mob (\vecr / a_p)$ when $a_p \gg A_s$, leading to $c_0 \sim a_p^{-1}$.
We can capture both regimes with the piecewise function
\begin{eqnarray}
    c_0 \approx 24v_{s}^2 \phi_s \left\lbrace      
        \begin{array}{ll}
            1 & \mathcal{R} < 1 \\
            \mathcal{R}^{-1} & \mathcal{R} \geqslant 1
        \end{array}
    \right. ,
\end{eqnarray}
where the front factor is read straight from our data, accounting for the known entrainment contribution.

We now turn our attention to the as yet undetermined time scales $\Tloop$ and $\TPeloop$.
$\Tloop$ is a representative time scale for the loop trajectories, which comes from an average over impact parameters.
We do not know its dependence on $a_p$, but we do know it is independent of temperature, as confirmed by Fig.~\ref{Fig: Example Cvv}(b) where the data at different temperature change sign at the same time.
Appendix \ref{Sec: Analysis of TPe} describes how to use this fact to isolate $\TPeloop$ in measurements of initial decay rates knowing only $\Trot$.
This approach finds
\begin{equation}
    \TPeloop \approx 1.73 \Perel a_s / v_s,
\end{equation}
proportional to $\Perel$ as expected, and associated with diffusion over a distance close to $A_s$.

Before progressing, we note that this decorrelation appears as a simple exponential because Brownian motion outside the entrained volume lacks the asymmetry that provided the square root in the analogous factor in the entrainment term.

Finally, we can feed $\TPeloop$ back into our fitted initial decay rates and solve for $\Tloop$.
This leads to Fig.~\ref{Fig: anasatz params}(b), where, similarly to $\Cvvp(0)$, we find it to be constant below $\mathcal{R} \lesssim 1$, while it rises with an apparent power law $\sim a_{p}^{1/2}$ above.

While the small $\mathcal{R}$ behaviour has the usual explanation that the passive particles are behaving as infinitesimal tracers, the apparent power law is harder to understand as it disagrees with the scaling argument in Appendix \ref{Sec: Scaling of Cvv and Tloop}, which predicts $\Tloop \sim a_p^{1}$.
This discrepancy could come from the periodicity of our system, which was not accounted for in our calculations, or could simply be a sign our data does not extend to large enough $a_p$ to see the expected behaviour.
For the purposes of this work it is sufficient to write down the empirical form as
\begin{eqnarray}
    \Tloop \approx 2 \frac{a_s}{v_s} \left\lbrace      
        \begin{array}{ll}
            1 & \mathcal{R} < 1 \\
            \mathcal{R}^{1/2} & \mathcal{R} \geqslant 1
        \end{array}
    \right. .
\end{eqnarray}
With this, our ansatz form for $\Cvv^{(p)}(t)$ is fully defined up to the single remaining free parameter $c_2$.


\section{Active diffusion}
\label{Sec: Active diff results}

\begin{figure}
    \includegraphics[width=8.5cm]{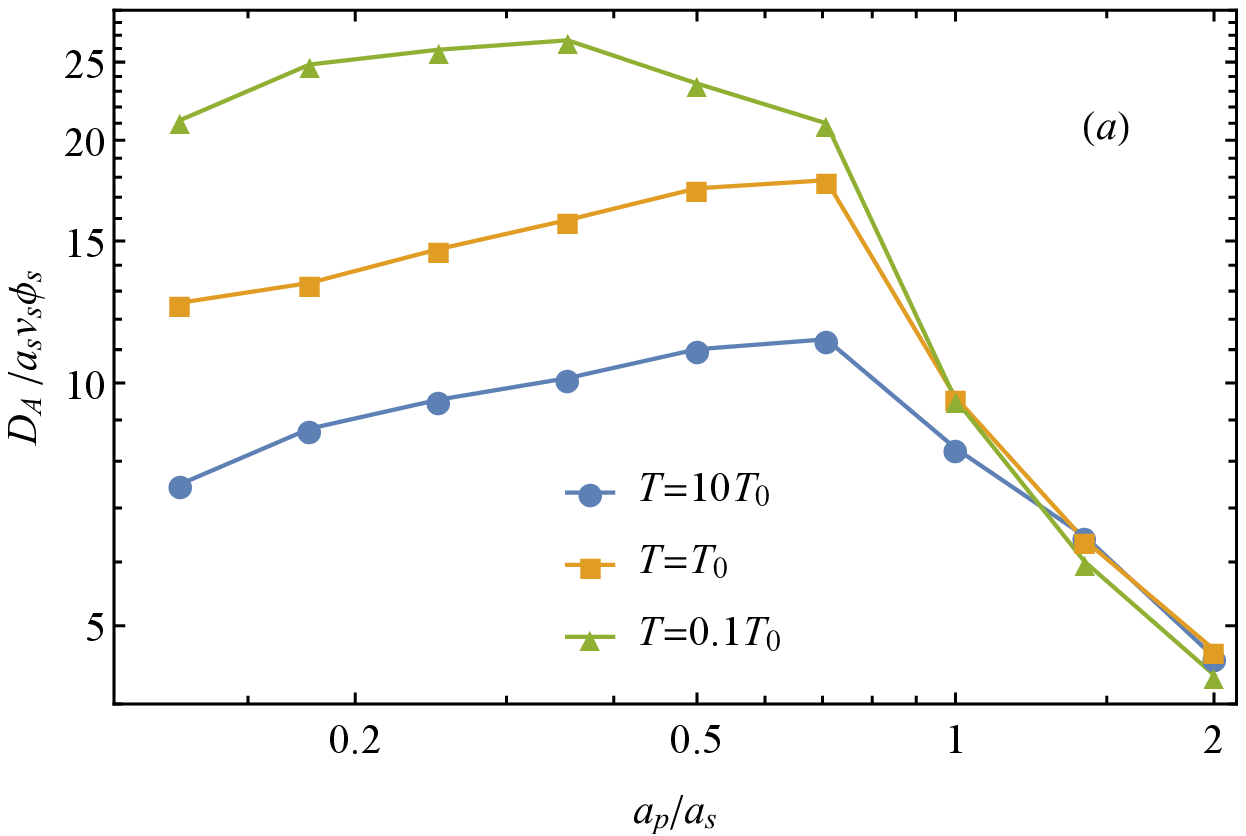}
    \includegraphics[width=8.5cm]{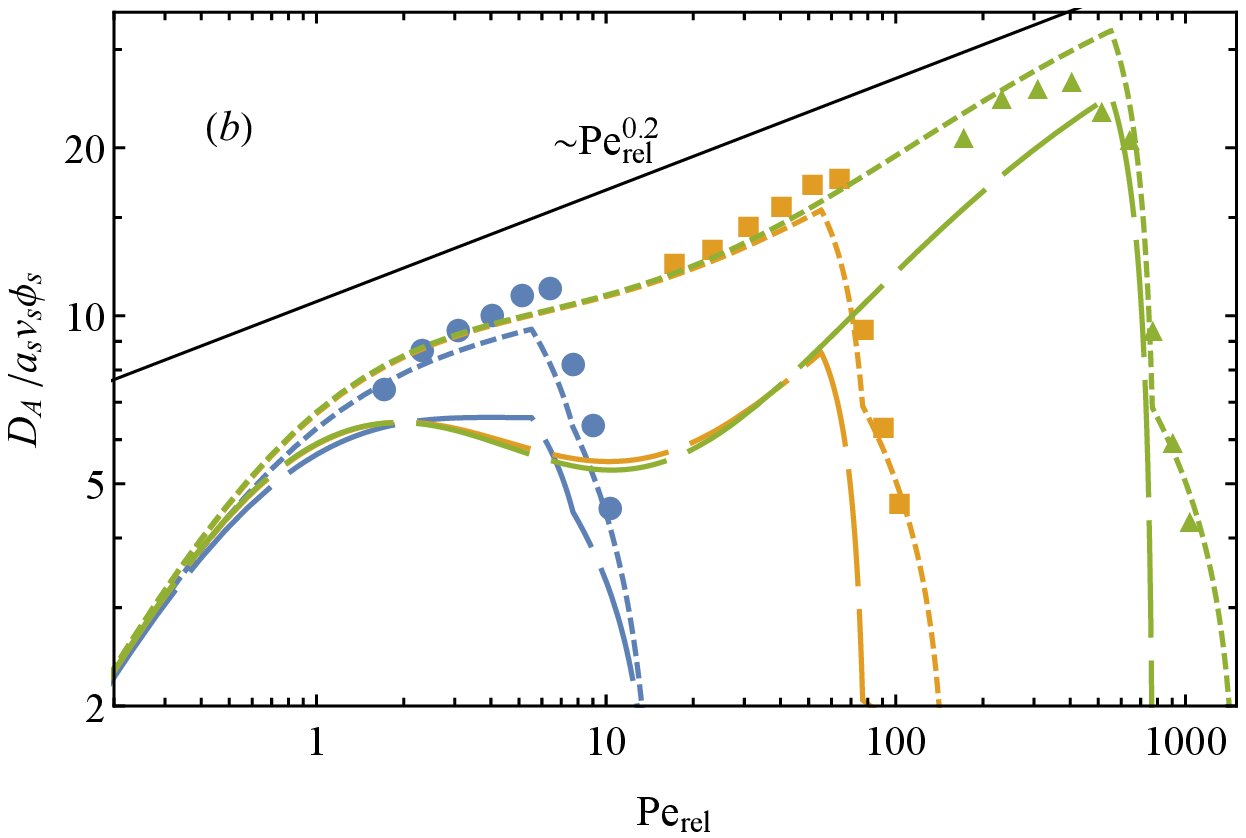}
    \caption{
    (a): active diffusivity calculated from simulation data, plotted against particle size.
    Data points are joined here to highlight the non-monotonic behaviour.
    Note the $T=T_0$ data is the same as the $N_s = 64$ data in Fig.~\ref{Fig: Dactive propto phi_s}.
    (b): the same data plotted against $\Perel$ and accompanied by plots of $D_A$ calculated with our ansatz function using $c_2 = 0.22$ (short dashes) and $c_2=0.5$ (long dashes).
    The solid line above the data is included to indicate the apparent power law we find across our $a_p \leqslant a_s /\sqrt{2} $ data.}
    \label{Fig: DA}
\end{figure}

We now move from the velocity autocorrelation to the active diffusivity, obtained by integrating $\Cvvp$ as per Eq.~\eqref{Eq: Green-Kubo relation}.
This we show for both numerical integration of the simulation data and analytical integration of our ansatz $\Cvvp$ using the approximate expressions for parameters in the previous section.
These are shown in Fig.~\ref{Fig: DA}.

Beginning with our data plotted against $a_p/a_s$ in Fig.~\ref{Fig: DA}(a), we observe two main features: non-monotonicity with a turning point just below $a_p/a_s = 1$; and a temperature-dependence on the small $a_p$ side.
The non-monotonicity requires different physics to be dominating at different regimes.
Using the understanding from previous sections, the presence and absence of entrainment at small and large $a_p$ respectively accounts for this behaviour, and is supported by the turning point being close to $\mathcal{R}=1$.

The quantitative behaviour in the two regimes can also be understood using our analysis of $\Cvvp$.
The decay on the large $a_p$ side comes primarily from the decay of $\Cvvp(0)$, which is not fully compensated for by the increase in $\Tloop$, at least not over the range of our data.
Note that $\Tloop \ll \Trt \ll \Trot, \TPeloop$ in the $T=0.1T_0$ data here, meaning the attenuation of the negative tail in $\Cvvp(t)$ is dominated by $\Tloop$, allowing us to neglect the other decorrelation times in this regime.

Work by Pushkin and Yeomans\cite{Pushkin2013PRL} has argued that the contribution from far field loops truncated by run and tumble events leads to a constant value of $D_A$ independent of the run length.
We expect our additional decorrelation mechanisms to fall under the same formalism, and hence would expect a constant term in $D_A$ that might be seen if we extended our range of $a_p/a_s$.
However, feeding our parameters into their calculation would put the value of this constant at $D_A / a_s v_s \phi_s = 20.25$, which is clearly missing or greatly reduced in our data.
We believe our periodic boundaries are the cause of its absence since it is an effect dominated by flow fields at impact parameters of order the run length.
In our case, the run length equals the side length of our box, so there will be significant interference from the swimmer's periodic images.

Our final comment on the large $a_p$ regime is that the collapse of the curves at different temperatures here is misleading.
As Fig.~\ref{Fig: Example Cvv}(b) showed, there is a significant difference in $\Cvvp$ here and the dynamics truly are affected by the temperature.
We believe our $10T_0$ data happened to have an equal loss of the negative tail and initial positive decay, but this is not generally true and we expect an intermediate temperature, e.g.~$5T_0$, would have a higher $D_A$ here because it's negative tail will have been affected the most by the decorrelations.
By the same reasoning, a temperature larger than $10T_0$ would have smaller $D_A$ because there is very little of the negative tail left to lose, leading to a greater loss from the positive part.

The variation at small $a_p$ is driven primarily by diffusive processes, so our $\Cvvp$ analysis predicts this behaviour to be a function of $\Trot \sim \Pe_s \sim T_0 / T$ and $\TPeent, \TPeloop \sim \Perel$, instead of a function of $a_p/a_s$.
In practice we find $\Trot$ is large enough that it has a negligible influence, leading to our data falling onto a master curve when plotted against $\Perel$, as shown in Fig.~\ref{Fig: DA}(b).
The apparent power law we observe across our data is close to $\Perel^{0.2}$, not $\Perel^{1}$ or $\Perel^{1/2}$ as we might have expected from the form of the two Brownian decorrelation factors.

Our ansatz function provides and explanation for this, although we must first specify a value of $c_2$.
The first value we use in Fig.~\ref{Fig: DA}(b) is $c_2 = 0.22$, which was chosen by a least squares fit of the ansatz to the data, setting all other parameters as described in Section \ref{Sec: Ansatz function}.
Here we see good agreement with the data at all temperatures, including the appearance of a shallow apparent power law.
Extending the plot down to smaller $\Perel$ finds the expected $\Perel^{1}$ behaviour does appear eventually.
Importantly, the ansatz plots have undulations, which are made extreme when using the largest allowed value of $c_2 = 0.5$.
This results from the two terms in the ansatz, with the loop term providing the peak at small $\Perel$ and the entrainment providing the second rise.
These undulations are more subtle in our simulation data, but they are still visible in the curvatures of the $10T_0$ and $1T_0$ data.
Hence we attribute the small power law to a transitional regime between loop and entrainment dominance.

It is useful here to compare to Kasyap, Koch and Wu's calculation of $D_{A}\sim \Pe^{1/2}$ for small $\Pe$ in a slender-body swimmer model \cite{Kasyap2014}.
In contrast, both terms in our anstaz lead to $D_{A} \sim \Pe^{1}$ in this limit.
We suspect the origin of this discrepancy might lie in their result assuming the distance swum in a single run is much smaller than the typical displacement by Brownian motion in the same time, whereas the reverse was true for all systems we used to construct our ansatz.

Finally, we note that the properties of our swimmers, especially the lack of steric interactions, will limit the applicability of our understanding to experimental systems.
Our nono-monotonicity is nevertheless in qualitative agreement with the experiments of Patteson et al.~\cite{Patteson2016}
This encourages us to propose that the cause is to be found in the transition between a regime dominated by entrainment events for small passive particles, and one for larger particles where far field loops are most important. 
We believe that this prediction could be testable with the experimental trajectories already available from the experiments in Ref.~\citenum{Patteson2016}.


\section{Conclusions}

We have looked at the effect of both particle size and temperature on the active diffusion of spherical passive particles in 3D periodic systems of microswimmers.
For this we used a `smart' version of the Wavelet Monte Carlo dynamics algorithm to simulate active systems with hydrodynamically correlated rotations and translations, biased by swimming force.
This gave us an efficient algorithm that includes correlated thermally-driven Brownian motion that is sensitive to particle size.

Our first results were geared towards validating active, non-thermal behaviour in smart WMCD, for which we simulated the trajectories of single passive particles at varying impact parameters from a passing swimmer at very large P\'{e}clet number.
These results were consistent with previous work, demonstrating the expected cusped-loop trajectories at large impact parameter, whose net and maximum displacements decayed with the expected power laws.

We then turned our attention to dilute mixtures of swimmers and passive particles with thermal fluctuations present.
By using a range of temperatures and passive particle sizes we were able to identify the physics driving active diffusion via the behaviour of the velocity autocorrelation.
Analysis of this led to constructing an ansatz function to unify the diverse forms of $\Cvvp$ observed.
This function was expressed as the sum of contributions from entrainment and non-entrained loop trajectories, both subject to exponential decorrelation factors coming from swimmer rotations and Brownian motion.
More generally, any mechanism causing passive particles to cross swimmer-induced stream lines could be included in this way.

Most parameters in our ansatz fall under one of two categories:
decorrelation times that vary with the appropriate P\'{e}clet number; and parameters describing the $\Pe\rightarrow \infty$ limit governed by the comparison between the hydrodynamic response of the passive particle and the geometric size of the swimmer.
By itself, the second category of parameters leads to $D_A$ having two regimes when plotted against $a_p$, with a decay away from the flat, small-$a_p$ regime where particles act as infinitesimal tracers.
The decorrelation factors then introduce a gradient to the small-$a_p$ regime, leading to non-monotonic behaviour.

Plotting $D_A$ against $\Perel$ reveals a master curve for the small $a_p$ regime.
The behaviour of this master curve over the range of P\'{e}clet numbers studied is made complicated by the entrainment and loop contributions appearing and plateauing at different values, with their sum leading to a weak apparent power law.

Finally, we note that the simplicity of our swimmer model means it is not expected to give quantitatively relevant results for comparison with experiment.
Instead, the strength of our results lies in the identification of the role of particle size and P\'{e}clet number(s) in the velocity autocorrelation.
In the process we highlighted the importance of temperature and near-field effects, both of which are often neglected in theoretical and computational studies of similar systems.

\begin{acknowledgments}
We gratefully acknowledge funding by the EPSRC, grant no.~EP/M508184/1, and the Warwick SCRTP for computational resources.
We also thank M.~Polin for several discussions and his help with preparing the manuscript. 
\end{acknowledgments}

\section*{Data Availability}
The data that support the findings of this study are available from the corresponding author upon reasonable request.

\appendix

\section{Loop size calculations}
\label{Sec: Loop size appendix}

We first consider the noise-free transits as graphed in Fig.~\ref{Fig: Particle loops}.  
The displacement of a passive particle $\vecdisp(t)$ due to the passage of a swimmer incident with impact parameter $\rho$ and swimmer velocity $v_s \hat{\vv{z}}$ obeys
\begin{equation}
d\vecdisp /dt= \vv v (\vecdisp + \rho \hat{\bm{\rho}}- v_{s} t \hat{\vv{z}} ) ,   
\end{equation}
where $\vv v(\vv r)$ is the flow field established by the swimmer and $\hat{\bm{\rho}}$ is the radial unit vector in cylindrical polar coordinates.

As constructed, the swimmer approaches from below so that $\disp_z-z_s$ decreases from $+\infty$, down to $-\infty$ for an orbit which does not get entrained. 
It is then convenient to write $\vv v(\vv r)=v_s \vv g(\rho \hat{\bm{\rho}}, z-z_s)$ and $v_s dt = dz_s$ so $z_s$ is the vertical rise of the swimmer, which leads to  
\begin{equation}
\vecdisp = \int\limits_{-\infty}^{z_s} \vv g(\rho \hat{\bm{\rho}} + \vecdisp_\rho, \disp_z - z_s') dz_s'.
\end{equation}

For the unentrained trajectories we can follow earlier work \cite{Pushkin2013JFluidMech} in expanding this for the total deflection as a series in $\vv g$ .  
It is convenient then to parameterise in terms of $z=-z_s$ which is to zero'th order the height of the passive above the swimmer, giving
\begin{equation}
\vecdisp = \int\limits_{z}^{\infty} 
    \vv g(\rho \hat{\bm{\rho}} + \vecdisp_\rho, \disp_z + z') dz'. 
\end{equation}
Expanding $\vecdisp$ in implied powers of $\vv g$  then gives $\vecdisp = \vecdisp^{(1)} + \vecdisp^{(2)} + O(\vv g^3)$ where 

\begin{equation}
\vecdisp^{(1)}(\rho\hat{\bm{\rho}},z) = \int\limits_{z}^{\infty}\vv g(\rho\hat{\bm{\rho}},z')dz'
\end{equation}
and
\begin{equation}
\vecdisp^{(2)}(\rho\hat{\bm{\rho}},z) = \int\limits_z^{\infty} \vecdisp^{(1)}(z') \cdot \vv\nabla \vv g(\rho\hat{\bm{\rho}},z')dz'.
\end{equation}
On the LHS $\rho$ and $z$ parameterise the transit in terms of impact parameter and time (as $-z/v_s$) through it.
However on the RHS they are simply cylindrical polar coordinates of the flow around the swimmer. 

We now focus on swimmers with azimuthal symmetry, so we write
\begin{equation}
\vecdisp^{(1)}(\rho \hat{\bm{\rho}},z) =\vv h(\rho, z),
\end{equation}
which can be thought of as a divergence free flow field.  
Moreover, in the far field the swimmer flow is proportional to that of a force dipole $-\partial(\Mobtt_{\mathrm{O}} \cdot \zhat)/\partial z$  so we infer that in the far field $\vv h$ approaches $\Mobtt_{\mathrm{O}} \cdot \zhat$, where $\Mobtt_{\mathrm{O}}$  is the Oseen tensor.
It then follows that for any force free swimmer the first order advective deflections of a point passive particle form a closed loop ending up with $\vecdisp^{(1)}(\rho, -\infty) = 0$.

At second order we now need
\begin{equation}
\vecdisp^{(2)}(\rho,z)= \int\limits_z^\infty  \vv h(\rho, z') \cdot \vv\nabla \left(-\frac{\partial}{\partial z'} \vv  h(\rho,z') \right) dz'.    
\end{equation}
As $\vv h$ is divergence free we can rewrite the integrand as $\vv\nabla \cdot \left[ \vv h(\rho, z') \left(-\frac{\partial}{\partial z'} \vv  h(\rho,z') \right) \right]$ leading to 
\begin{multline}
\vecdisp^{(2)}(\rho,z) = 
    h_z(\rho, z) \frac{\partial}{\partial z} \vv  h(\rho,z)  \\ - \frac{1}{\rho}\frac{\partial}{\partial \rho}\rho 
    \int\limits_z^{\infty}   h_\rho(\rho, z')  \frac{\partial}{\partial z'} \vv h(\rho,z')dz'.
\end{multline}
For the transverse displacement we can now give a full result
\begin{equation}
\disp^{(2)}_\rho(\rho,z) = 
    h_z(\rho, z) \frac{\partial}{\partial z}   h_\rho(\rho,z) + \frac{1}{2\rho}\frac{\partial}{\partial \rho}\rho \,h_\rho(\rho, z)^2,
\end{equation}
and note that this gives zero for the complete transit.  

The longitudinal displacement is given by 
\begin{multline} 
\disp^{(2)}_z(\rho,z) = 
    h_z(\rho, z) \frac{\partial}{\partial z}   h_z(\rho,z) \\ - \frac{1}{\rho}\frac{\partial}{\partial \rho}\rho 
    \int\limits_z^{\infty}   h_\rho(\rho, z')  \frac{\partial}{\partial z'} h_z(\rho,z')dz'.
\end{multline}
The integrand here can be expressed using the divergence free property of $\vv h$ as
\begin{equation}
h_\rho(\rho, z') \left( -\frac{1}{\rho}\frac{\partial}{\partial \rho} \rho  h_\rho(\rho,z') \right) 
    = -\left( \frac{1}{\rho} +\frac{1}{2}\frac{\partial}{\partial \rho} \right)  h_\rho(\rho, z')^2 .    
\end{equation}
This then gives the deflection through a full transit as 
\begin{equation}
\disp^{(2)}_z(\rho,-\infty) =
  \frac{1}{\rho^2} \rho \frac{ \partial}{\partial \rho} 
  \left(1 +\frac{1}{2}\rho\frac{\partial}{\partial \rho}\right)
  \int\limits_{-\infty}^\infty   h_\rho(\rho, z')^2 dz'. 
\end{equation}

In the far field force dipole limit $\vv h$  is just given by the Oseen tensor and we have $\int_{-\infty}^\infty   h_\rho(\rho, z')^2 dz' \propto +1/\rho $ and hence $\disp_z^{(2)}\propto - 1/\rho^3$,  where the negative sign signifies dominance by back-flow around the swimmer.  
For a swimmer modelled as two opposed point forces separated by $A_s$ and with dipole strength $\kappa = A_s F_s$, near approaches  $\rho \ll  A_s$ lead to $\int_{-\infty}^\infty   h_\rho(\rho, z')^2 dz' \propto +\rho/A_{s}^{2} $ and hence $\disp_z^{(2)}\propto + 1/(\rho A_{s}^{2})$, this time with a positive sign.

The full forms for a swimmer modelled by two point forces separated by $A_s$ can be found as follows.  
The transverse component of the Oseen tensor is given by $ (\rho z / 8\pi\eta) (\rho^2+z^2)^{-3/2} $ and integrating this with respect to $z$ gives $-(\rho /8\pi\eta) (\rho^2+z^2)^{-1/2} $, which then leads to 
\begin{multline}
h_\rho(\rho, z) =
\frac{\kappa}{8\pi\eta A_s} \left[ (1 + (\zeta - \alpha)^2)^{-1/2} \right. 
    \\- \left. (1 + (\zeta + \alpha)^2)^{-1/2} \right]
\end{multline}
where $\zeta = z/\rho$ and $\alpha=A_s/2\rho$.  
In the limit of large $\alpha$  the integral over  $h_\rho(\rho, z')^2$ is then dominated by two well separated Lorentzians each of width $\rho$ and height $\propto A_{s}^{-2}$, leading to $\int_{-\infty}^\infty  h_\rho(\rho, z')^2 dz'  \propto \rho/A_{s}^{2}$  as used in the paragraph above. 
 
To get the full result we write 
\begin{equation}
\int\limits_{-\infty}^\infty   h_\rho(\rho, z')^2 dz' 
 = \frac{\rho \kappa^2}{(8\pi\eta A_s)^2} [I(0)-2I(\alpha)+I(0)],
\end{equation}
where 
\begin{equation}\begin{aligned} 
I(\alpha) & =  \int\limits_{-\infty}^\infty
    \left(1+(\zeta-\alpha)^2\right)^{-1/2} \left(1+(\zeta + \alpha)^2 \right)^{-1/2}  d\zeta \\
& =\frac{1}{\pi}\int\limits_{-\infty}^\infty  
    \int\limits_{-\infty}^\infty \int\limits_{-\infty}^\infty 
        d\zeta \, d\lambda \, d\mu \Big[ 
        \exp\left(-\lambda^2 (1+(\zeta-\alpha)^2)\right)
            \\& \hphantom{some more space} \times 
            \exp\left(-\mu^2 (1+(\zeta+\alpha)^2)\right) \Big] \\
& =\frac{1}{\sqrt{\pi}}
    \int\limits_{-\infty}^\infty \int\limits_{-\infty}^\infty  
        d\lambda \,d\mu \, \Bigg[ (\lambda^2+\mu^2)^{-1/2}
        \\ & \hphantom{some space} \times 
        \exp\left( -(\lambda^2+\mu^2) -\frac{4\alpha^2\lambda^2\mu^2}{\lambda^2+\mu^2}\right) \Bigg].
\end{aligned} \end{equation}
Next we change from $\lambda$, $\mu$ as a Cartesian coordinate pair to the equivalent plane polars to give
\begin{equation}\begin{aligned} 
I(\alpha) &=  \frac{1}{\sqrt{\pi}}
    \int\limits_0^\infty \int\limits_0^{2\pi}dr\, d\theta\, 
        \exp\left(-r^2(1+4\alpha^2 \sin^{2}\theta\, \cos^{2}\theta)\right) \\
& = \frac{1}{2}\int\limits_0^{2\pi} d\theta 
    (1+\alpha^2 \sin^2{2\theta})^{-1/2} = 2 K(i\alpha)
\end{aligned} \end{equation}
where $K(x)=\int_0^{\pi/2} d\theta (1-x^2 \sin^2{\theta})^{-1/2}$ is the complete elliptic integral of the first kind.


\section{Large $a_p$ behaviour of $\Cvv(0)$ and $\Tloop$}
\label{Sec: Scaling of Cvv and Tloop}

Here we use scaling arguments to calculate the expected behaviour of $\Cvv(0)$ and $\Tloop$ in the large $a_p$ limit.
We begin by considering the far field, where the swimmer looks like a dipole force that produces a flow field given by the stresslet \cite{Pushkin2013PRL}
\begin{eqnarray}
    \vecv_{d}(\vecr) & = &
    - \kappa \vhat_s \cdot (\vhat_s \cdot \nabla) \Mobtt_{\mathrm{O}}(\vecr)
    \nonumber
    \\
    & = & \frac{\kappa}{r^3}
        ( 3 ( \vhat_s \cdot \rhat )^2 - 1 )\vecr
\label{Eq: Stresslet}
\end{eqnarray}
where $\kappa = A_s F_s / 8\pi\eta$ includes the dipole strength and the numerical factors from the Oseen tensor $\Mobtt_{\mathrm{O}}$.

Although this is independent of $a_p$, it only applies at distances larger than order $a_p$ from the swimmer.
This boundary can be incorporated by expressing the velocity as
\begin{equation}
    \vecv_{d}(\vecr) = a_p^{-2} \vecv_0(\vecr / a_p).
\label{Eq: Velocity scaling relation}
\end{equation}
This scaling relation can be extended to the near field for particles with $a_p \gg A_s$ because the mobility tensor in WMCD - which replaces $\Mobtt_{\mathrm{O}}$ in Eq.~\eqref{Eq: Stresslet} - can be written as $\Mobtt_{\mathrm{WMCD}}(\vecr) = a_p^{-1} \Mob (\vecr/a_p)$.

The scaling of $\Cvv(0)$ is then easily calculated with
\begin{eqnarray}
    \Cvv(0) & = & a_p^{-4} \phi_s \int d^{3}\vecr \, \vecv_0(\vecr/a_p)^2 \nonumber \\
        & = & a_p^{-1} \phi_s \int d^{3}\vecr' \, \vecv_0 (\vecr')^2 .
\end{eqnarray}
The remaining integral is now just a constant for fixed swimmer parameters, so we have $\Cvv(0) \sim a_p^{-1}$.

To predict the behaviour of $\Tloop$ we can take the relative speed to be $v_s$ since the speed of the passive particle is everywhere less than $F_s / 6\pi\eta a_p \propto (a_s/a_p)v_s \ll v_s$.
Alongside the length scale $a_p$, this leads to the relevant time scale for changes in the swimmer flow field being $\Tloop = a_p / v_s$.


\section{Analysis of $\TPe$}
\label{Sec: Analysis of TPe}

In this section we demonstrate how to identify the behaviour of $\TPeloop$.
Using the ansatz form in Eq.~\eqref{Eq: Ansatz loop term} and ignoring the entrainment term on the grounds that $\phi_{\ent}\Cvv^{(s)}(0)  \ll c_0$ in our simulations, measuring the initial decay rate of $\Cvv$ gives the full decay rate
\begin{equation}
    \Ttot^{-1} = \Tloop^{-1} + \Trt^{-1} + \Trot^{-1} + \TPeloop^{-1}.
\end{equation}
Note we included the quadratic term in the fit, with $c_2/\Tloop^2$ as a second fitted variable, although we do not use those values in this work.

The different terms in $\Ttot$ are all expected to have different behaviours across our simulations: $\Trt$ is a constant; with fixed swimmer parameters, $\Trot$ depends only on the temperature; and $\Tloop$ depends on particle sizes but not the temperature.
By looking at the difference of $\Ttot$ at different temperatures but the same $a_p$, we can remove the influence of both $\Trt$ and $\Tloop$.
We then only need to remove $\Trot$, which can be done by hand since we know its form, leading us to consider
\begin{equation}\begin{aligned}
    \mathcal{T}^{-1} & = \Ttot(a_p,T_i)^{-1} - \Ttot(a_p,T_j)^{-1}
        - \frac{k_B (T_i - T_j)}{4\pi\eta a_s^3} 
\label{Eq: T definition}\\
        & =  \TPeloop(a_p,T_i)^{-1} - \TPeloop(a_p,T_j)^{-1}. 
\end{aligned}\end{equation}

\begin{figure}
    \includegraphics[width=8.5cm]{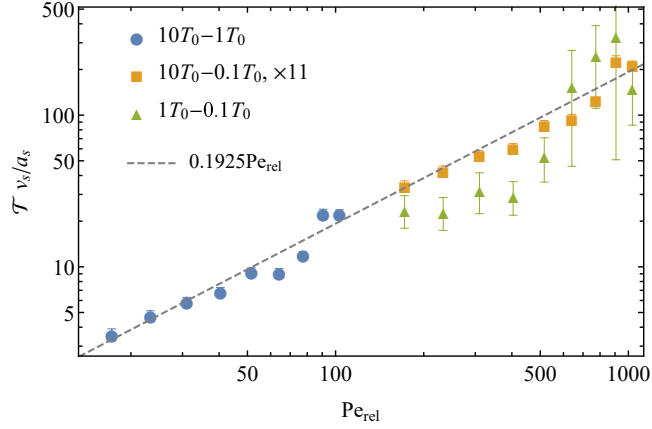}
    \caption{Plot of $\mathcal{T}$, defined in Eq.~\eqref{Eq: T definition}, against P\'{e}clet number for our 3 combinations of temperature.
    The $10T_0 - 0.1T_0$ data are multiplied by 11 to account for a their larger factor of $(T_i/T_j)^1 - 1$.
    Errors shown are 1 standard deviation, and are smaller than the plot markers for many data points.}
    \label{Fig: T vs Pe}
\end{figure}

Assuming a power law $\TPeloop(a_p,T) \propto \Perel^{\alpha}$ we would have
\begin{equation}
    \mathcal{T} \propto \frac{\Perel(a_p,T_j)^{\alpha} }{ \left(T_i / T_j\right)^{\alpha} - 1},
\end{equation}
so we should still see the same power law if we are consistent with which temperature we use in $\Perel$.
We have 3 combinations of temperature: $10T_0$ and $1T_0$; $10T_0$ and $0.1T_0$; and $1T_0$ and $0.1T_0$.
For all combinations we use the smaller temperature in $\Perel$, giving the results in Fig.~\ref{Fig: T vs Pe}.
Here we see $\mathcal{T} \sim \Perel^{1}$, so hence so does $\TPeloop$, which is consistent with other time scales increasing linearly with a P\'{e}clet number.
Furthermore, Fig.~\ref{Fig: T vs Pe} provides us with the constant of proportionality, which is 9 times larger than the one for $\mathcal{T}$ due to the factor of $(T_i/T_j)^1 - 1$.
Hence, we have
\begin{eqnarray}
    \TPeloop(\Perel) = 1.73 \Perel a_s/v_s.
\end{eqnarray}

\bibliography{Active_diff}

\begin{thebibliography}{50}%
\makeatletter
\providecommand \@ifxundefined [1]{%
 \@ifx{#1\undefined}
}%
\providecommand \@ifnum [1]{%
 \ifnum #1\expandafter \@firstoftwo
 \else \expandafter \@secondoftwo
 \fi
}%
\providecommand \@ifx [1]{%
 \ifx #1\expandafter \@firstoftwo
 \else \expandafter \@secondoftwo
 \fi
}%
\providecommand \natexlab [1]{#1}%
\providecommand \enquote  [1]{``#1''}%
\providecommand \bibnamefont  [1]{#1}%
\providecommand \bibfnamefont [1]{#1}%
\providecommand \citenamefont [1]{#1}%
\providecommand \href@noop [0]{\@secondoftwo}%
\providecommand \href [0]{\begingroup \@sanitize@url \@href}%
\providecommand \@href[1]{\@@startlink{#1}\@@href}%
\providecommand \@@href[1]{\endgroup#1\@@endlink}%
\providecommand \@sanitize@url [0]{\catcode `\\12\catcode `\$12\catcode
  `\&12\catcode `\#12\catcode `\^12\catcode `\_12\catcode `\%12\relax}%
\providecommand \@@startlink[1]{}%
\providecommand \@@endlink[0]{}%
\providecommand \url  [0]{\begingroup\@sanitize@url \@url }%
\providecommand \@url [1]{\endgroup\@href {#1}{\urlprefix }}%
\providecommand \urlprefix  [0]{URL }%
\providecommand \Eprint [0]{\href }%
\providecommand \doibase [0]{http://dx.doi.org/}%
\providecommand \selectlanguage [0]{\@gobble}%
\providecommand \bibinfo  [0]{\@secondoftwo}%
\providecommand \bibfield  [0]{\@secondoftwo}%
\providecommand \translation [1]{[#1]}%
\providecommand \BibitemOpen [0]{}%
\providecommand \bibitemStop [0]{}%
\providecommand \bibitemNoStop [0]{.\EOS\space}%
\providecommand \EOS [0]{\spacefactor3000\relax}%
\providecommand \BibitemShut  [1]{\csname bibitem#1\endcsname}%
\let\auto@bib@innerbib\@empty
\bibitem [{\citenamefont {Wu}\ and\ \citenamefont {Libchaber}(2000)}]{Wu2000}%
  \BibitemOpen
  \bibfield  {author} {\bibinfo {author} {\bibfnamefont {X.-L.}\ \bibnamefont
  {Wu}}\ and\ \bibinfo {author} {\bibfnamefont {A.}~\bibnamefont {Libchaber}},\
  }\bibfield  {title} {\enquote {\bibinfo {title} {Particle diffusion in a
  quasi-two-dimensional bacterial bath},}\ }\href@noop {} {\bibfield  {journal}
  {\bibinfo  {journal} {Phys. Rev. Lett.}\ }\textbf {\bibinfo {volume} {84}},\
  \bibinfo {pages} {3017--3020} (\bibinfo {year} {2000})}\BibitemShut {NoStop}%
\bibitem [{\citenamefont {Soni}\ \emph {et~al.}(2003)\citenamefont {Soni},
  \citenamefont {Ali}, \citenamefont {Hatwalne},\ and\ \citenamefont
  {Shivashankar}}]{Soni2003}%
  \BibitemOpen
  \bibfield  {author} {\bibinfo {author} {\bibfnamefont {G.~V.}\ \bibnamefont
  {Soni}}, \bibinfo {author} {\bibfnamefont {B.~M.~J.}\ \bibnamefont {Ali}},
  \bibinfo {author} {\bibfnamefont {Y.}~\bibnamefont {Hatwalne}}, \ and\
  \bibinfo {author} {\bibfnamefont {G.~V.}\ \bibnamefont {Shivashankar}},\
  }\bibfield  {title} {\enquote {\bibinfo {title} {Single particle tracking of
  correlated bacterial dynamics},}\ }\href@noop {} {\bibfield  {journal}
  {\bibinfo  {journal} {Biophys. J.}\ }\textbf {\bibinfo {volume} {84}},\
  \bibinfo {pages} {2634--2637} (\bibinfo {year} {2003})}\BibitemShut {NoStop}%
\bibitem [{\citenamefont {Leptos}\ \emph {et~al.}(2009)\citenamefont {Leptos},
  \citenamefont {Guasto}, \citenamefont {Gollub}, \citenamefont {Pesci},\ and\
  \citenamefont {Goldstein}}]{Leptos2009}%
  \BibitemOpen
  \bibfield  {author} {\bibinfo {author} {\bibfnamefont {K.~C.}\ \bibnamefont
  {Leptos}}, \bibinfo {author} {\bibfnamefont {J.~S.}\ \bibnamefont {Guasto}},
  \bibinfo {author} {\bibfnamefont {J.~P.}\ \bibnamefont {Gollub}}, \bibinfo
  {author} {\bibfnamefont {A.~I.}\ \bibnamefont {Pesci}}, \ and\ \bibinfo
  {author} {\bibfnamefont {R.~E.}\ \bibnamefont {Goldstein}},\ }\bibfield
  {title} {\enquote {\bibinfo {title} {Dynamics of enhanced tracer diffusion in
  suspensions of swimming eukaryotic microorganisms},}\ }\href@noop {}
  {\bibfield  {journal} {\bibinfo  {journal} {Phys. Rev. Lett.}\ }\textbf
  {\bibinfo {volume} {103}},\ \bibinfo {pages} {198103} (\bibinfo {year}
  {2009})}\BibitemShut {NoStop}%
\bibitem [{\citenamefont {Mi{\~n}o}\ \emph {et~al.}(2011)\citenamefont
  {Mi{\~n}o}, \citenamefont {Mallouk}, \citenamefont {Darnige}, \citenamefont
  {Hoyos}, \citenamefont {Dauchet}, \citenamefont {Dunstan}, \citenamefont
  {Soto}, \citenamefont {Wang}, \citenamefont {Rousselet},\ and\ \citenamefont
  {Clement}}]{Mino2011}%
  \BibitemOpen
  \bibfield  {author} {\bibinfo {author} {\bibfnamefont {G.}~\bibnamefont
  {Mi{\~n}o}}, \bibinfo {author} {\bibfnamefont {T.~E.}\ \bibnamefont
  {Mallouk}}, \bibinfo {author} {\bibfnamefont {T.}~\bibnamefont {Darnige}},
  \bibinfo {author} {\bibfnamefont {M.}~\bibnamefont {Hoyos}}, \bibinfo
  {author} {\bibfnamefont {J.}~\bibnamefont {Dauchet}}, \bibinfo {author}
  {\bibfnamefont {J.}~\bibnamefont {Dunstan}}, \bibinfo {author} {\bibfnamefont
  {R.}~\bibnamefont {Soto}}, \bibinfo {author} {\bibfnamefont {Y.}~\bibnamefont
  {Wang}}, \bibinfo {author} {\bibfnamefont {A.}~\bibnamefont {Rousselet}}, \
  and\ \bibinfo {author} {\bibfnamefont {E.}~\bibnamefont {Clement}},\
  }\bibfield  {title} {\enquote {\bibinfo {title} {Enhanced diffusion due to
  active swimmers at a solid surface},}\ }\href@noop {} {\bibfield  {journal}
  {\bibinfo  {journal} {Phys. Rev. Lett.}\ }\textbf {\bibinfo {volume} {106}},\
  \bibinfo {pages} {048102} (\bibinfo {year} {2011})}\BibitemShut {NoStop}%
\bibitem [{\citenamefont {Kurtuldu}\ \emph {et~al.}(2011)\citenamefont
  {Kurtuldu}, \citenamefont {Guasto}, \citenamefont {Johnson},\ and\
  \citenamefont {Gollub}}]{Kurtuldu2011}%
  \BibitemOpen
  \bibfield  {author} {\bibinfo {author} {\bibfnamefont {H.}~\bibnamefont
  {Kurtuldu}}, \bibinfo {author} {\bibfnamefont {J.~S.}\ \bibnamefont
  {Guasto}}, \bibinfo {author} {\bibfnamefont {K.~A.}\ \bibnamefont {Johnson}},
  \ and\ \bibinfo {author} {\bibfnamefont {J.~P.}\ \bibnamefont {Gollub}},\
  }\bibfield  {title} {\enquote {\bibinfo {title} {Enhancement of biomixing by
  swimming algal cells in two-dimensional films},}\ }\href@noop {} {\bibfield
  {journal} {\bibinfo  {journal} {Proceedings of the National Academy of
  Sciences}\ }\textbf {\bibinfo {volume} {108}},\ \bibinfo {pages}
  {10391--10395} (\bibinfo {year} {2011})}\BibitemShut {NoStop}%
\bibitem [{\citenamefont {Valeriani}\ \emph {et~al.}(2011)\citenamefont
  {Valeriani}, \citenamefont {Li}, \citenamefont {Novosel}, \citenamefont
  {Arlt},\ and\ \citenamefont {Marenduzzo}}]{Valeriani2011}%
  \BibitemOpen
  \bibfield  {author} {\bibinfo {author} {\bibfnamefont {C.}~\bibnamefont
  {Valeriani}}, \bibinfo {author} {\bibfnamefont {M.}~\bibnamefont {Li}},
  \bibinfo {author} {\bibfnamefont {J.}~\bibnamefont {Novosel}}, \bibinfo
  {author} {\bibfnamefont {J.}~\bibnamefont {Arlt}}, \ and\ \bibinfo {author}
  {\bibfnamefont {D.}~\bibnamefont {Marenduzzo}},\ }\bibfield  {title}
  {\enquote {\bibinfo {title} {Colloids in a bacterial bath: simulations and
  experiments},}\ }\href@noop {} {\bibfield  {journal} {\bibinfo  {journal}
  {Soft Matter}\ }\textbf {\bibinfo {volume} {7}},\ \bibinfo {pages}
  {5228--5238} (\bibinfo {year} {2011})}\BibitemShut {NoStop}%
\bibitem [{\citenamefont {Mi{\~n}o}\ \emph {et~al.}(2013)\citenamefont
  {Mi{\~n}o}, \citenamefont {Dunstan}, \citenamefont {Rousselet}, \citenamefont
  {Cl{\'e}ment},\ and\ \citenamefont {Soto}}]{Mino2013}%
  \BibitemOpen
  \bibfield  {author} {\bibinfo {author} {\bibfnamefont {G.~L.}\ \bibnamefont
  {Mi{\~n}o}}, \bibinfo {author} {\bibfnamefont {J.}~\bibnamefont {Dunstan}},
  \bibinfo {author} {\bibfnamefont {A.}~\bibnamefont {Rousselet}}, \bibinfo
  {author} {\bibfnamefont {E.}~\bibnamefont {Cl{\'e}ment}}, \ and\ \bibinfo
  {author} {\bibfnamefont {R.}~\bibnamefont {Soto}},\ }\bibfield  {title}
  {\enquote {\bibinfo {title} {Induced diffusion of tracers in a bacterial
  suspension: theory and experiments},}\ }\href@noop {} {\bibfield  {journal}
  {\bibinfo  {journal} {Journal of Fluid Mechanics}\ }\textbf {\bibinfo
  {volume} {729}},\ \bibinfo {pages} {423--444} (\bibinfo {year}
  {2013})}\BibitemShut {NoStop}%
\bibitem [{\citenamefont {Jepson}\ \emph {et~al.}(2013)\citenamefont {Jepson},
  \citenamefont {Martinez}, \citenamefont {Schwarz-Linek}, \citenamefont
  {Morozov},\ and\ \citenamefont {Poon}}]{Jepson2013}%
  \BibitemOpen
  \bibfield  {author} {\bibinfo {author} {\bibfnamefont {A.}~\bibnamefont
  {Jepson}}, \bibinfo {author} {\bibfnamefont {V.~A.}\ \bibnamefont
  {Martinez}}, \bibinfo {author} {\bibfnamefont {J.}~\bibnamefont
  {Schwarz-Linek}}, \bibinfo {author} {\bibfnamefont {A.}~\bibnamefont
  {Morozov}}, \ and\ \bibinfo {author} {\bibfnamefont {W.~C.~K.}\ \bibnamefont
  {Poon}},\ }\bibfield  {title} {\enquote {\bibinfo {title} {Enhanced diffusion
  of nonswimmers in a three-dimensional bath of motile bacteria},}\ }\href@noop
  {} {\bibfield  {journal} {\bibinfo  {journal} {Phys. Rev. E}\ }\textbf
  {\bibinfo {volume} {88}},\ \bibinfo {pages} {041002} (\bibinfo {year}
  {2013})}\BibitemShut {NoStop}%
\bibitem [{\citenamefont {Jeanneret}\ \emph {et~al.}(2016)\citenamefont
  {Jeanneret}, \citenamefont {Pushkin}, \citenamefont {Kantsler},\ and\
  \citenamefont {Polin}}]{Jeanneret2016}%
  \BibitemOpen
  \bibfield  {author} {\bibinfo {author} {\bibfnamefont {R.}~\bibnamefont
  {Jeanneret}}, \bibinfo {author} {\bibfnamefont {D.~O.}\ \bibnamefont
  {Pushkin}}, \bibinfo {author} {\bibfnamefont {V.}~\bibnamefont {Kantsler}}, \
  and\ \bibinfo {author} {\bibfnamefont {M.}~\bibnamefont {Polin}},\ }\bibfield
   {title} {\enquote {\bibinfo {title} {Entrainment dominates the interaction
  of microalgae with micron-sized objects},}\ }\href@noop {} {\bibfield
  {journal} {\bibinfo  {journal} {Nature Communications}\ }\textbf {\bibinfo
  {volume} {7}},\ \bibinfo {pages} {12518} (\bibinfo {year}
  {2016})}\BibitemShut {NoStop}%
\bibitem [{\citenamefont {Mathijssen}, \citenamefont {Jeanneret},\ and\
  \citenamefont {Polin}(2018)}]{Mathijssen2018}%
  \BibitemOpen
  \bibfield  {author} {\bibinfo {author} {\bibfnamefont {A.~J. T.~M.}\
  \bibnamefont {Mathijssen}}, \bibinfo {author} {\bibfnamefont
  {R.}~\bibnamefont {Jeanneret}}, \ and\ \bibinfo {author} {\bibfnamefont
  {M.}~\bibnamefont {Polin}},\ }\bibfield  {title} {\enquote {\bibinfo {title}
  {Universal entrainment mechanism controls contact times with motile cells},}\
  }\href@noop {} {\bibfield  {journal} {\bibinfo  {journal} {Phys. Rev.
  Fluids}\ }\textbf {\bibinfo {volume} {3}},\ \bibinfo {pages} {033103}
  (\bibinfo {year} {2018})}\BibitemShut {NoStop}%
\bibitem [{\citenamefont {Underhill}, \citenamefont {Hernandez-Ortiz},\ and\
  \citenamefont {Graham}(2008)}]{Underhill2008}%
  \BibitemOpen
  \bibfield  {author} {\bibinfo {author} {\bibfnamefont {P.~T.}\ \bibnamefont
  {Underhill}}, \bibinfo {author} {\bibfnamefont {J.~P.}\ \bibnamefont
  {Hernandez-Ortiz}}, \ and\ \bibinfo {author} {\bibfnamefont {M.~D.}\
  \bibnamefont {Graham}},\ }\bibfield  {title} {\enquote {\bibinfo {title}
  {Diffusion and spatial correlations in suspensions of swimming particles},}\
  }\href@noop {} {\bibfield  {journal} {\bibinfo  {journal} {Phys. Rev. Lett.}\
  }\textbf {\bibinfo {volume} {100}},\ \bibinfo {pages} {248101} (\bibinfo
  {year} {2008})}\BibitemShut {NoStop}%
\bibitem [{\citenamefont {Molina}\ and\ \citenamefont
  {Yamamoto}(2014)}]{Molina2014}%
  \BibitemOpen
  \bibfield  {author} {\bibinfo {author} {\bibfnamefont {J.~J.}\ \bibnamefont
  {Molina}}\ and\ \bibinfo {author} {\bibfnamefont {R.}~\bibnamefont
  {Yamamoto}},\ }\bibfield  {title} {\enquote {\bibinfo {title} {Diffusion of
  colloidal particles in swimming suspensions},}\ }\href@noop {} {\bibfield
  {journal} {\bibinfo  {journal} {Molecular Physics}\ }\textbf {\bibinfo
  {volume} {112}},\ \bibinfo {pages} {1389--1397} (\bibinfo {year}
  {2014})}\BibitemShut {NoStop}%
\bibitem [{\citenamefont {Morozov}\ and\ \citenamefont
  {Marenduzzo}(2014)}]{Morozov2014}%
  \BibitemOpen
  \bibfield  {author} {\bibinfo {author} {\bibfnamefont {A.}~\bibnamefont
  {Morozov}}\ and\ \bibinfo {author} {\bibfnamefont {D.}~\bibnamefont
  {Marenduzzo}},\ }\bibfield  {title} {\enquote {\bibinfo {title} {Enhanced
  diffusion of tracer particles in dilute bacterial suspensions},}\ }\href@noop
  {} {\bibfield  {journal} {\bibinfo  {journal} {Soft Matter}\ }\textbf
  {\bibinfo {volume} {10}},\ \bibinfo {pages} {2748--2758} (\bibinfo {year}
  {2014})}\BibitemShut {NoStop}%
\bibitem [{\citenamefont {Krafnick}\ and\ \citenamefont
  {Garc{\'{\i}}a}(2015)}]{Krafnick2015}%
  \BibitemOpen
  \bibfield  {author} {\bibinfo {author} {\bibfnamefont {R.~C.}\ \bibnamefont
  {Krafnick}}\ and\ \bibinfo {author} {\bibfnamefont {A.~E.}\ \bibnamefont
  {Garc{\'{\i}}a}},\ }\bibfield  {title} {\enquote {\bibinfo {title} {Impact of
  hydrodynamics on effective interactions in suspensions of active and passive
  matter},}\ }\href@noop {} {\bibfield  {journal} {\bibinfo  {journal} {Phys.
  Rev. E}\ }\textbf {\bibinfo {volume} {91}},\ \bibinfo {pages} {022308}
  (\bibinfo {year} {2015})}\BibitemShut {NoStop}%
\bibitem [{\citenamefont {Krishnamurthy}\ and\ \citenamefont
  {Subramanian}(2015)}]{Krishnamurthy2015}%
  \BibitemOpen
  \bibfield  {author} {\bibinfo {author} {\bibfnamefont {D.}~\bibnamefont
  {Krishnamurthy}}\ and\ \bibinfo {author} {\bibfnamefont {G.}~\bibnamefont
  {Subramanian}},\ }\bibfield  {title} {\enquote {\bibinfo {title} {Collective
  motion in a suspension of micro-swimmers that run-and-tumble and rotary
  diffuse},}\ }\href@noop {} {\bibfield  {journal} {\bibinfo  {journal}
  {Journal of Fluid Mechanics}\ }\textbf {\bibinfo {volume} {781}},\ \bibinfo
  {pages} {422--466} (\bibinfo {year} {2015})}\BibitemShut {NoStop}%
\bibitem [{\citenamefont {de~Graaf}\ and\ \citenamefont
  {Stenhammar}(2017)}]{deGraaf2017}%
  \BibitemOpen
  \bibfield  {author} {\bibinfo {author} {\bibfnamefont {J.}~\bibnamefont
  {de~Graaf}}\ and\ \bibinfo {author} {\bibfnamefont {J.}~\bibnamefont
  {Stenhammar}},\ }\bibfield  {title} {\enquote {\bibinfo {title}
  {Lattice-{B}oltzmann simulations of microswimmer-tracer interactions},}\
  }\href@noop {} {\bibfield  {journal} {\bibinfo  {journal} {Phys. Rev. E}\
  }\textbf {\bibinfo {volume} {95}},\ \bibinfo {pages} {023302} (\bibinfo
  {year} {2017})}\BibitemShut {NoStop}%
\bibitem [{\citenamefont {Shum}\ and\ \citenamefont
  {Yeomans}(2017)}]{Shum2017}%
  \BibitemOpen
  \bibfield  {author} {\bibinfo {author} {\bibfnamefont {H.}~\bibnamefont
  {Shum}}\ and\ \bibinfo {author} {\bibfnamefont {J.~M.}\ \bibnamefont
  {Yeomans}},\ }\bibfield  {title} {\enquote {\bibinfo {title} {Entrainment and
  scattering in microswimmer-colloid interactions},}\ }\href@noop {} {\bibfield
   {journal} {\bibinfo  {journal} {Phys. Rev. Fluids}\ }\textbf {\bibinfo
  {volume} {2}},\ \bibinfo {pages} {113101} (\bibinfo {year}
  {2017})}\BibitemShut {NoStop}%
\bibitem [{\citenamefont {Harder}\ and\ \citenamefont
  {Cacciuto}(2018)}]{Harder2018}%
  \BibitemOpen
  \bibfield  {author} {\bibinfo {author} {\bibfnamefont {J.}~\bibnamefont
  {Harder}}\ and\ \bibinfo {author} {\bibfnamefont {A.}~\bibnamefont
  {Cacciuto}},\ }\bibfield  {title} {\enquote {\bibinfo {title} {Hierarchical
  collective motion of a mixture of active dipolar {J}anus particles and
  passive charged colloids in two dimensions},}\ }\href@noop {} {\bibfield
  {journal} {\bibinfo  {journal} {Phys. Rev. E}\ }\textbf {\bibinfo {volume}
  {97}},\ \bibinfo {pages} {022603} (\bibinfo {year} {2018})}\BibitemShut
  {NoStop}%
\bibitem [{\citenamefont {Dunkel}\ \emph {et~al.}(2010)\citenamefont {Dunkel},
  \citenamefont {Putz}, \citenamefont {Zaid},\ and\ \citenamefont
  {Yeomans}}]{Dunkel2010}%
  \BibitemOpen
  \bibfield  {author} {\bibinfo {author} {\bibfnamefont {J.}~\bibnamefont
  {Dunkel}}, \bibinfo {author} {\bibfnamefont {V.~B.}\ \bibnamefont {Putz}},
  \bibinfo {author} {\bibfnamefont {I.~M.}\ \bibnamefont {Zaid}}, \ and\
  \bibinfo {author} {\bibfnamefont {J.~M.}\ \bibnamefont {Yeomans}},\
  }\bibfield  {title} {\enquote {\bibinfo {title} {Swimmer-tracer scattering at
  low {R}eynolds number},}\ }\href@noop {} {\bibfield  {journal} {\bibinfo
  {journal} {Soft Matter}\ }\textbf {\bibinfo {volume} {6}},\ \bibinfo {pages}
  {4268--4276} (\bibinfo {year} {2010})}\BibitemShut {NoStop}%
\bibitem [{\citenamefont {Thiffeault}\ and\ \citenamefont
  {Childress}(2010)}]{Thiffeault2010}%
  \BibitemOpen
  \bibfield  {author} {\bibinfo {author} {\bibfnamefont {J.-L.}\ \bibnamefont
  {Thiffeault}}\ and\ \bibinfo {author} {\bibfnamefont {S.}~\bibnamefont
  {Childress}},\ }\bibfield  {title} {\enquote {\bibinfo {title} {Stirring by
  swimming bodies},}\ }\href@noop {} {\bibfield  {journal} {\bibinfo  {journal}
  {Physics Letters A}\ }\textbf {\bibinfo {volume} {374}},\ \bibinfo {pages}
  {3487 -- 3490} (\bibinfo {year} {2010})}\BibitemShut {NoStop}%
\bibitem [{\citenamefont {Eckhardt}\ and\ \citenamefont
  {Zammert}(2012)}]{Eckhardt2012}%
  \BibitemOpen
  \bibfield  {author} {\bibinfo {author} {\bibfnamefont {B.}~\bibnamefont
  {Eckhardt}}\ and\ \bibinfo {author} {\bibfnamefont {S.}~\bibnamefont
  {Zammert}},\ }\bibfield  {title} {\enquote {\bibinfo {title} {Non-normal
  tracer diffusion from stirring by swimming microorganisms},}\ }\href@noop {}
  {\bibfield  {journal} {\bibinfo  {journal} {Eur. Phys. J. E}\ }\textbf
  {\bibinfo {volume} {35}},\ \bibinfo {pages} {96} (\bibinfo {year}
  {2012})}\BibitemShut {NoStop}%
\bibitem [{\citenamefont {Mathijssen}, \citenamefont {Pushkin},\ and\
  \citenamefont {Yeomans}(2015)}]{Mathijssen2015}%
  \BibitemOpen
  \bibfield  {author} {\bibinfo {author} {\bibfnamefont {A.~J. T.~M.}\
  \bibnamefont {Mathijssen}}, \bibinfo {author} {\bibfnamefont {D.~O.}\
  \bibnamefont {Pushkin}}, \ and\ \bibinfo {author} {\bibfnamefont {J.~M.}\
  \bibnamefont {Yeomans}},\ }\bibfield  {title} {\enquote {\bibinfo {title}
  {Tracer trajectories and displacement due to a micro-swimmer near a
  surface},}\ }\href@noop {} {\bibfield  {journal} {\bibinfo  {journal}
  {Journal of Fluid Mechanics}\ }\textbf {\bibinfo {volume} {773}},\ \bibinfo
  {pages} {498--519} (\bibinfo {year} {2015})}\BibitemShut {NoStop}%
\bibitem [{\citenamefont {Thiffeault}(2015)}]{Thiffeault2015}%
  \BibitemOpen
  \bibfield  {author} {\bibinfo {author} {\bibfnamefont {J.-L.}\ \bibnamefont
  {Thiffeault}},\ }\bibfield  {title} {\enquote {\bibinfo {title} {Distribution
  of particle displacements due to swimming microorganisms},}\ }\href@noop {}
  {\bibfield  {journal} {\bibinfo  {journal} {Phys. Rev. E}\ }\textbf {\bibinfo
  {volume} {92}},\ \bibinfo {pages} {023023} (\bibinfo {year}
  {2015})}\BibitemShut {NoStop}%
\bibitem [{\citenamefont {Suma}, \citenamefont {Cugliandolo},\ and\
  \citenamefont {Gonnella}(2016)}]{Suma2016}%
  \BibitemOpen
  \bibfield  {author} {\bibinfo {author} {\bibfnamefont {A.}~\bibnamefont
  {Suma}}, \bibinfo {author} {\bibfnamefont {L.~F.}\ \bibnamefont
  {Cugliandolo}}, \ and\ \bibinfo {author} {\bibfnamefont {G.}~\bibnamefont
  {Gonnella}},\ }\bibfield  {title} {\enquote {\bibinfo {title} {Tracer motion
  in an active dumbbell fluid},}\ }\href@noop {} {\bibfield  {journal}
  {\bibinfo  {journal} {Journal of Statistical Mechanics: Theory and
  Experiment}\ }\textbf {\bibinfo {volume} {2016}},\ \bibinfo {pages} {054029}
  (\bibinfo {year} {2016})}\BibitemShut {NoStop}%
\bibitem [{\citenamefont {Burkholder}\ and\ \citenamefont
  {Brady}(2017)}]{Burkholder2017}%
  \BibitemOpen
  \bibfield  {author} {\bibinfo {author} {\bibfnamefont {E.~W.}\ \bibnamefont
  {Burkholder}}\ and\ \bibinfo {author} {\bibfnamefont {J.~F.}\ \bibnamefont
  {Brady}},\ }\bibfield  {title} {\enquote {\bibinfo {title} {Tracer diffusion
  in active suspensions},}\ }\href@noop {} {\bibfield  {journal} {\bibinfo
  {journal} {Phys. Rev. E}\ }\textbf {\bibinfo {volume} {95}},\ \bibinfo
  {pages} {052605} (\bibinfo {year} {2017})}\BibitemShut {NoStop}%
\bibitem [{\citenamefont {Yasuda}, \citenamefont {Okamoto},\ and\ \citenamefont
  {Komura}(2017)}]{Yasuda2017}%
  \BibitemOpen
  \bibfield  {author} {\bibinfo {author} {\bibfnamefont {K.}~\bibnamefont
  {Yasuda}}, \bibinfo {author} {\bibfnamefont {R.}~\bibnamefont {Okamoto}}, \
  and\ \bibinfo {author} {\bibfnamefont {S.}~\bibnamefont {Komura}},\
  }\bibfield  {title} {\enquote {\bibinfo {title} {Anomalous diffusion in
  viscoelastic media with active force dipoles},}\ }\href@noop {} {\bibfield
  {journal} {\bibinfo  {journal} {Phys. Rev. E}\ }\textbf {\bibinfo {volume}
  {95}},\ \bibinfo {pages} {032417} (\bibinfo {year} {2017})}\BibitemShut
  {NoStop}%
\bibitem [{\citenamefont {Mueller}\ and\ \citenamefont
  {Thiffeault}(2017)}]{Mueller2017}%
  \BibitemOpen
  \bibfield  {author} {\bibinfo {author} {\bibfnamefont {P.}~\bibnamefont
  {Mueller}}\ and\ \bibinfo {author} {\bibfnamefont {J.-L.}\ \bibnamefont
  {Thiffeault}},\ }\bibfield  {title} {\enquote {\bibinfo {title} {Fluid
  transport and mixing by an unsteady microswimmer},}\ }\href@noop {}
  {\bibfield  {journal} {\bibinfo  {journal} {Phys. Rev. Fluids}\ }\textbf
  {\bibinfo {volume} {2}},\ \bibinfo {pages} {013103} (\bibinfo {year}
  {2017})}\BibitemShut {NoStop}%
\bibitem [{\citenamefont {Fax\'{e}n}(1922)}]{Faxen1922}%
  \BibitemOpen
  \bibfield  {author} {\bibinfo {author} {\bibfnamefont {H.}~\bibnamefont
  {Fax\'{e}n}},\ }\bibfield  {title} {\enquote {\bibinfo {title} {{Der
  Widerstand gegen die Bewegung einer starren Kugel in einer z\"{a}hen
  Fl\"{u}ssigkeit, die zwischen zwei parallelen ebenen W\"{a}nden
  eingeschlossen ist}},}\ }\href@noop {} {\bibfield  {journal} {\bibinfo
  {journal} {Annalen der Physik}\ }\textbf {\bibinfo {volume} {373}},\ \bibinfo
  {pages} {89--119} (\bibinfo {year} {1922})}\BibitemShut {NoStop}%
\bibitem [{\citenamefont {Durlofsky}, \citenamefont {Brady},\ and\
  \citenamefont {Bossis}(1987)}]{DurlofskyBradyBossis1987}%
  \BibitemOpen
  \bibfield  {author} {\bibinfo {author} {\bibfnamefont {L.}~\bibnamefont
  {Durlofsky}}, \bibinfo {author} {\bibfnamefont {J.~F.}\ \bibnamefont
  {Brady}}, \ and\ \bibinfo {author} {\bibfnamefont {G.}~\bibnamefont
  {Bossis}},\ }\bibfield  {title} {\enquote {\bibinfo {title} {Dynamic
  simulation of hydrodynamically interacting particles},}\ }\href@noop {}
  {\bibfield  {journal} {\bibinfo  {journal} {Journal of Fluid Mechanics}\
  }\textbf {\bibinfo {volume} {180}},\ \bibinfo {pages} {21--49} (\bibinfo
  {year} {1987})}\BibitemShut {NoStop}%
\bibitem [{\citenamefont {Rotne}\ and\ \citenamefont
  {Prager}(1969)}]{RotnePrager1969}%
  \BibitemOpen
  \bibfield  {author} {\bibinfo {author} {\bibfnamefont {J.}~\bibnamefont
  {Rotne}}\ and\ \bibinfo {author} {\bibfnamefont {S.}~\bibnamefont {Prager}},\
  }\bibfield  {title} {\enquote {\bibinfo {title} {Variational treatment of
  hydrodynamic interaction in polymers},}\ }\href@noop {} {\bibfield  {journal}
  {\bibinfo  {journal} {The Journal of Chemical Physics}\ }\textbf {\bibinfo
  {volume} {50}},\ \bibinfo {pages} {4831--4837} (\bibinfo {year}
  {1969})}\BibitemShut {NoStop}%
\bibitem [{\citenamefont {Yamakawa}(1970)}]{Yamakawa1970}%
  \BibitemOpen
  \bibfield  {author} {\bibinfo {author} {\bibfnamefont {H.}~\bibnamefont
  {Yamakawa}},\ }\bibfield  {title} {\enquote {\bibinfo {title} {Transport
  properties of polymer chains in dilute solution: Hydrodynamic interaction},}\
  }\href@noop {} {\bibfield  {journal} {\bibinfo  {journal} {The Journal of
  Chemical Physics}\ }\textbf {\bibinfo {volume} {53}},\ \bibinfo {pages}
  {436--443} (\bibinfo {year} {1970})}\BibitemShut {NoStop}%
\bibitem [{\citenamefont {Kasyap}, \citenamefont {Koch},\ and\ \citenamefont
  {Wu}(2014)}]{Kasyap2014}%
  \BibitemOpen
  \bibfield  {author} {\bibinfo {author} {\bibfnamefont {T.~V.}\ \bibnamefont
  {Kasyap}}, \bibinfo {author} {\bibfnamefont {D.~L.}\ \bibnamefont {Koch}}, \
  and\ \bibinfo {author} {\bibfnamefont {M.}~\bibnamefont {Wu}},\ }\bibfield
  {title} {\enquote {\bibinfo {title} {Hydrodynamic tracer diffusion in
  suspensions of swimming bacteria},}\ }\href@noop {} {\bibfield  {journal}
  {\bibinfo  {journal} {Physics of Fluids}\ }\textbf {\bibinfo {volume} {26}},\
  \bibinfo {pages} {081901} (\bibinfo {year} {2014})}\BibitemShut {NoStop}%
\bibitem [{\citenamefont {Patteson}\ \emph {et~al.}(2016)\citenamefont
  {Patteson}, \citenamefont {Gopinath}, \citenamefont {Purohit},\ and\
  \citenamefont {Arratia}}]{Patteson2016}%
  \BibitemOpen
  \bibfield  {author} {\bibinfo {author} {\bibfnamefont {A.~E.}\ \bibnamefont
  {Patteson}}, \bibinfo {author} {\bibfnamefont {A.}~\bibnamefont {Gopinath}},
  \bibinfo {author} {\bibfnamefont {P.~K.}\ \bibnamefont {Purohit}}, \ and\
  \bibinfo {author} {\bibfnamefont {P.~E.}\ \bibnamefont {Arratia}},\
  }\bibfield  {title} {\enquote {\bibinfo {title} {Particle diffusion in active
  fluids is non-monotonic in size},}\ }\href@noop {} {\bibfield  {journal}
  {\bibinfo  {journal} {Soft Matter}\ }\textbf {\bibinfo {volume} {12}},\
  \bibinfo {pages} {2365--2372} (\bibinfo {year} {2016})}\BibitemShut {NoStop}%
\bibitem [{\citenamefont {Taylor}(1953)}]{Taylor1953}%
  \BibitemOpen
  \bibfield  {author} {\bibinfo {author} {\bibfnamefont {G.~I.}\ \bibnamefont
  {Taylor}},\ }\bibfield  {title} {\enquote {\bibinfo {title} {Dispersion of
  soluble matter in solvent flowing slowly through a tube},}\ }\href@noop {}
  {\bibfield  {journal} {\bibinfo  {journal} {Proceedings of the Royal Society
  of London. Series A. Mathematical and Physical Sciences}\ }\textbf {\bibinfo
  {volume} {219}},\ \bibinfo {pages} {186--203} (\bibinfo {year}
  {1953})}\BibitemShut {NoStop}%
\bibitem [{\citenamefont {Pham}\ \emph {et~al.}(2009)\citenamefont {Pham},
  \citenamefont {Schiller}, \citenamefont {Prakash},\ and\ \citenamefont
  {D{\"u}nweg}}]{Pham2009}%
  \BibitemOpen
  \bibfield  {author} {\bibinfo {author} {\bibfnamefont {T.~T.}\ \bibnamefont
  {Pham}}, \bibinfo {author} {\bibfnamefont {U.~D.}\ \bibnamefont {Schiller}},
  \bibinfo {author} {\bibfnamefont {J.~R.}\ \bibnamefont {Prakash}}, \ and\
  \bibinfo {author} {\bibfnamefont {B.}~\bibnamefont {D{\"u}nweg}},\ }\bibfield
   {title} {\enquote {\bibinfo {title} {Implicit and explicit solvent models
  for the simulation of a single polymer chain in solution: Lattice {B}oltzmann
  versus {B}rownian dynamics},}\ }\href@noop {} {\bibfield  {journal} {\bibinfo
   {journal} {The Journal of Chemical Physics}\ }\textbf {\bibinfo {volume}
  {131}},\ \bibinfo {pages} {164114} (\bibinfo {year} {2009})}\BibitemShut
  {NoStop}%
\bibitem [{\citenamefont {Jain}\ \emph {et~al.}(2012)\citenamefont {Jain},
  \citenamefont {P.}, \citenamefont {D{\"u}nweg},\ and\ \citenamefont
  {Prakash}}]{Jain2012}%
  \BibitemOpen
  \bibfield  {author} {\bibinfo {author} {\bibfnamefont {A.}~\bibnamefont
  {Jain}}, \bibinfo {author} {\bibfnamefont {S.}~\bibnamefont {P.}}, \bibinfo
  {author} {\bibfnamefont {B.}~\bibnamefont {D{\"u}nweg}}, \ and\ \bibinfo
  {author} {\bibfnamefont {J.~R.}\ \bibnamefont {Prakash}},\ }\bibfield
  {title} {\enquote {\bibinfo {title} {Optimization of a {B}rownian-dynamics
  algorithm for semidilute polymer solutions},}\ }\href@noop {} {\bibfield
  {journal} {\bibinfo  {journal} {Physical Review E}\ }\textbf {\bibinfo
  {volume} {85}},\ \bibinfo {pages} {066703} (\bibinfo {year}
  {2012})}\BibitemShut {NoStop}%
\bibitem [{\citenamefont {Dyer}\ and\ \citenamefont
  {Ball}(2017)}]{DyerBall2017}%
  \BibitemOpen
  \bibfield  {author} {\bibinfo {author} {\bibfnamefont {O.~T.}\ \bibnamefont
  {Dyer}}\ and\ \bibinfo {author} {\bibfnamefont {R.~C.}\ \bibnamefont
  {Ball}},\ }\bibfield  {title} {\enquote {\bibinfo {title} {Wavelet {M}onte
  {C}arlo dynamics: A new algorithm for simulating the hydrodynamics of
  interacting {B}rownian particles},}\ }\href@noop {} {\bibfield  {journal}
  {\bibinfo  {journal} {The Journal of Chemical Physics}\ }\textbf {\bibinfo
  {volume} {146}},\ \bibinfo {pages} {124111} (\bibinfo {year}
  {2017})}\BibitemShut {NoStop}%
\bibitem [{\citenamefont {Dyer}(2019)}]{Dyer2019}%
  \BibitemOpen
  \bibfield  {author} {\bibinfo {author} {\bibfnamefont {O.~T.}\ \bibnamefont
  {Dyer}},\ }\emph {\bibinfo {title} {Wavelet Monte Carlo dynamics}},\
  \href@noop {} {Ph.D. thesis},\ \bibinfo  {school} {University of Warwick}
  (\bibinfo {year} {2019})\BibitemShut {NoStop}%
\bibitem [{\citenamefont {D{\"u}nweg}\ and\ \citenamefont
  {Kremer}(1993)}]{Dunweg1993}%
  \BibitemOpen
  \bibfield  {author} {\bibinfo {author} {\bibfnamefont {B.}~\bibnamefont
  {D{\"u}nweg}}\ and\ \bibinfo {author} {\bibfnamefont {K.}~\bibnamefont
  {Kremer}},\ }\bibfield  {title} {\enquote {\bibinfo {title} {Molecular
  dynamics simulation of a polymer chain in solution},}\ }\href@noop {}
  {\bibfield  {journal} {\bibinfo  {journal} {The Journal of Chemical Physics}\
  }\textbf {\bibinfo {volume} {99}},\ \bibinfo {pages} {6983--6997} (\bibinfo
  {year} {1993})}\BibitemShut {NoStop}%
\bibitem [{\citenamefont {Pushkin}\ and\ \citenamefont
  {Yeomans}(2013)}]{Pushkin2013PRL}%
  \BibitemOpen
  \bibfield  {author} {\bibinfo {author} {\bibfnamefont {D.~O.}\ \bibnamefont
  {Pushkin}}\ and\ \bibinfo {author} {\bibfnamefont {J.~M.}\ \bibnamefont
  {Yeomans}},\ }\bibfield  {title} {\enquote {\bibinfo {title} {Fluid mixing by
  curved trajectories of microswimmers},}\ }\href@noop {} {\bibfield  {journal}
  {\bibinfo  {journal} {Phys. Rev. Lett.}\ }\textbf {\bibinfo {volume} {111}},\
  \bibinfo {pages} {188101} (\bibinfo {year} {2013})}\BibitemShut {NoStop}%
\bibitem [{\citenamefont {Green}(1954)}]{Green1954}%
  \BibitemOpen
  \bibfield  {author} {\bibinfo {author} {\bibfnamefont {M.~S.}\ \bibnamefont
  {Green}},\ }\bibfield  {title} {\enquote {\bibinfo {title} {Markoff random
  processes and the statistical mechanics of time-dependent phenomena. {II}.
  {I}rreversible processes in fluids},}\ }\href@noop {} {\bibfield  {journal}
  {\bibinfo  {journal} {The Journal of Chemical Physics}\ }\textbf {\bibinfo
  {volume} {22}},\ \bibinfo {pages} {398--413} (\bibinfo {year}
  {1954})}\BibitemShut {NoStop}%
\bibitem [{\citenamefont {Kubo}(1957)}]{Kubo1957}%
  \BibitemOpen
  \bibfield  {author} {\bibinfo {author} {\bibfnamefont {R.}~\bibnamefont
  {Kubo}},\ }\bibfield  {title} {\enquote {\bibinfo {title}
  {Statistical-mechanical theory of irreversible processes. {I}. {G}eneral
  theory and simple applications to magnetic and conduction problems},}\
  }\href@noop {} {\bibfield  {journal} {\bibinfo  {journal} {Journal of the
  Physical Society of Japan}\ }\textbf {\bibinfo {volume} {12}},\ \bibinfo
  {pages} {570--586} (\bibinfo {year} {1957})}\BibitemShut {NoStop}%
\bibitem [{\citenamefont {Happel}\ and\ \citenamefont
  {Brenner}(1973)}]{Happel1973}%
  \BibitemOpen
  \bibfield  {author} {\bibinfo {author} {\bibfnamefont {J.}~\bibnamefont
  {Happel}}\ and\ \bibinfo {author} {\bibfnamefont {H.}~\bibnamefont
  {Brenner}},\ }\href@noop {} {\emph {\bibinfo {title} {Low {R}eynolds Number
  Hydrodynamics}}}\ (\bibinfo  {publisher} {Noordhoff,Leyden},\ \bibinfo {year}
  {1973})\BibitemShut {NoStop}%
\bibitem [{\citenamefont {Kubo}(1966)}]{Kubo1966}%
  \BibitemOpen
  \bibfield  {author} {\bibinfo {author} {\bibfnamefont {R.}~\bibnamefont
  {Kubo}},\ }\bibfield  {title} {\enquote {\bibinfo {title} {The
  fluctuation-dissipation theorem},}\ }\href@noop {} {\bibfield  {journal}
  {\bibinfo  {journal} {Reports on Progress in Physics}\ }\textbf {\bibinfo
  {volume} {29}},\ \bibinfo {pages} {255--284} (\bibinfo {year}
  {1966})}\BibitemShut {NoStop}%
\bibitem [{\citenamefont {Noetinger}(1990)}]{Noetinger1990}%
  \BibitemOpen
  \bibfield  {author} {\bibinfo {author} {\bibfnamefont {B.}~\bibnamefont
  {Noetinger}},\ }\bibfield  {title} {\enquote {\bibinfo {title} {Fluctuating
  hydrodynamics and {B}rownian motion},}\ }\href@noop {} {\bibfield  {journal}
  {\bibinfo  {journal} {Physica A: Statistical Mechanics and its Applications}\
  }\textbf {\bibinfo {volume} {163}},\ \bibinfo {pages} {545 -- 558} (\bibinfo
  {year} {1990})}\BibitemShut {NoStop}%
\bibitem [{\citenamefont {Rossky}, \citenamefont {Doll},\ and\ \citenamefont
  {Friedman}(1978)}]{Rossky1978}%
  \BibitemOpen
  \bibfield  {author} {\bibinfo {author} {\bibfnamefont {P.~J.}\ \bibnamefont
  {Rossky}}, \bibinfo {author} {\bibfnamefont {J.~D.}\ \bibnamefont {Doll}}, \
  and\ \bibinfo {author} {\bibfnamefont {H.~L.}\ \bibnamefont {Friedman}},\
  }\bibfield  {title} {\enquote {\bibinfo {title} {Brownian dynamics as smart
  {M}onte {C}arlo simulation},}\ }\href@noop {} {\bibfield  {journal} {\bibinfo
   {journal} {The Journal of Chemical Physics}\ }\textbf {\bibinfo {volume}
  {69}},\ \bibinfo {pages} {4628--4633} (\bibinfo {year} {1978})}\BibitemShut
  {NoStop}%
\bibitem [{\citenamefont {Berg}(1993)}]{Berg1993}%
  \BibitemOpen
  \bibfield  {author} {\bibinfo {author} {\bibfnamefont {H.}~\bibnamefont
  {Berg}},\ }\href@noop {} {\emph {\bibinfo {title} {Random Walks in
  Biology}}},\ Princeton paperbacks\ (\bibinfo  {publisher} {Princeton
  University Press},\ \bibinfo {year} {1993})\BibitemShut {NoStop}%
\bibitem [{\citenamefont {Saragosti}, \citenamefont {Silberzan},\ and\
  \citenamefont {Buguin}(2012)}]{Saragosti2012}%
  \BibitemOpen
  \bibfield  {author} {\bibinfo {author} {\bibfnamefont {J.}~\bibnamefont
  {Saragosti}}, \bibinfo {author} {\bibfnamefont {P.}~\bibnamefont
  {Silberzan}}, \ and\ \bibinfo {author} {\bibfnamefont {A.}~\bibnamefont
  {Buguin}},\ }\bibfield  {title} {\enquote {\bibinfo {title} {Modeling {E}.
  coli tumbles by rotational diffusion. {I}mplications for chemotaxis},}\
  }\href@noop {} {\bibfield  {journal} {\bibinfo  {journal} {PLOS ONE}\
  }\textbf {\bibinfo {volume} {7}},\ \bibinfo {pages} {1--6} (\bibinfo {year}
  {2012})}\BibitemShut {NoStop}%
\bibitem [{\citenamefont {Pushkin}, \citenamefont {Shum},\ and\ \citenamefont
  {Yeomans}(2013)}]{Pushkin2013JFluidMech}%
  \BibitemOpen
  \bibfield  {author} {\bibinfo {author} {\bibfnamefont {D.~O.}\ \bibnamefont
  {Pushkin}}, \bibinfo {author} {\bibfnamefont {H.}~\bibnamefont {Shum}}, \
  and\ \bibinfo {author} {\bibfnamefont {J.~M.}\ \bibnamefont {Yeomans}},\
  }\bibfield  {title} {\enquote {\bibinfo {title} {Fluid transport by
  individual microswimmers},}\ }\href@noop {} {\bibfield  {journal} {\bibinfo
  {journal} {Journal of Fluid Mechanics}\ }\textbf {\bibinfo {volume} {726}},\
  \bibinfo {pages} {5--25} (\bibinfo {year} {2013})}\BibitemShut {NoStop}%
\bibitem [{\citenamefont {Crank}(1975)}]{Crank1975}%
  \BibitemOpen
  \bibfield  {author} {\bibinfo {author} {\bibfnamefont {J.}~\bibnamefont
  {Crank}},\ }\href@noop {} {\emph {\bibinfo {title} {The Mathematics of
  Diffusion}}},\ \bibinfo {edition} {2nd}\ ed.\ (\bibinfo  {publisher} {Oxford
  University Press},\ \bibinfo {year} {1975})\BibitemShut {NoStop}%
\end{thebibliography}%

\end{document}